\documentclass[aps,prb,amsmath,amssymb,nobibnotes,longbibliography,superscriptaddress,twocolumn]{revtex4-2}

\usepackage{t1enc}
\usepackage[utf8]{inputenc}
\usepackage{amsmath}
\usepackage{dsfont}
\usepackage{ulem}
\usepackage{nicematrix}
\usepackage{tikz}
\usepackage{cancel}

\usepackage{graphicx,bm,amsmath,amssymb,natbib,color}
\usepackage[unicode=true, colorlinks=true, citecolor={blue!80!black}, urlcolor={blue!50!black}, linkcolor = {blue!80!black}]{hyperref}
\usepackage{physics}
\usepackage{bm}
\usepackage{bbm}
\usepackage{soul}

\usepackage[english]{babel}

\begin{document}
\title{Landau-Zener-Stückelberg spectroscopy of a fluxonium quantum circuit}
\newcommand{\affA}{\affiliation{Grupo de Circuitos Cuánticos Bariloche, Div. Dispositivos y Sensores, Centro Atómico Bariloche-CNEA, 8400 San Carlos de Bariloche, Argentina.}}
\newcommand{\affB}{\affiliation{Instituto Balseiro (Universidad Nacional de Cuyo), Bariloche, Argentina.}}
\newcommand{\affC}{\affiliation{Centro At\'omico Bariloche and Instituto Balseiro, 8400 San Carlos de Bariloche, R\'io Negro, Argentina.}}
\newcommand{\affD}{\affiliation{Instituto de Nanociencia y Nanotecnolog\'{\i}a (INN), CONICET-CNEA, Argentina.}}

\author{Valentín Reparaz}
\affA
\affB
\author{María José Sánchez}
\affC\affD
\author{Maximiliano Gatto}
\affA
\affB
\author{Daniel Dominguez}
\affC
\author{Leandro Tosi}
\email[Corresponding author:]{leandro.tosi@ib.edu.ar}
\affA
\affB
\date{\today}

\begin{abstract}
In this work, we study the time-averaged populations obtained for a fluxonium circuit under a large amplitude nonresonant periodic drive. We present numerical simulations of the time evolution which consider the multi-level structure of the driven quantum circuit, looking for a realistic modeling closer to experimental implementations. The Landau-Zener-Stückelberg spectra show resonances that can be understood as originated from constructive interference favoring transitions to higher levels. For a truncated two-level system (TLS) the resonance patterns can be interpreted using a simplified description of the avoided crossing that takes into account the dynamic phase accumulated at each operation point. For the multilevel case, we derive an effective two-level Hamiltonian using a Schrieffer-Wolff transformation starting from the Floquet Hamiltonian in the Sambe space. 
Our study provides predictive insight into experimental outcomes, offering an intuitive interpretation that could also support the implementation of fast-non-adiabatic single-qubit gates and entangling protocols.
   
\end{abstract}
\maketitle

\section{Introduction}
Superconducting circuits are among the most promising candidates for quantum computing, due to continuous improvements in their performances,  coherence times, processor size and  gate fidelities \cite{Krantz2019,  kjaergaard_2020}.  Although the transmon \cite{Koch2007} has become a widely established choice to encode quantum information, it has been argued that the fluxonium qubit introduced shortly after \cite{Manucharyan2009,Koch2009, Pop2014}, is a comparable or even better option for building quantum processors \cite{nguyen_2019,gyenis2021, bao_2022, weiss_2022, nguyen_2022,ding_2023}: It is  unaffected by charge noise \cite{Koch2009} and can be made fairly insensitive to flux noise \cite{lin_2018,nguyen_2019, nguyen_2022}. As its  scalability is comparable to that of the transmon, it can be easily coupled to do two-qubit operations \cite{nesterov_2021,moska_2021,nguyen_2022,moska_2022}. In addition, there are new technological advances, such as granular aluminum \cite{maleeva2018circuit,grunhaupt2019granular, levy2019electrodynamics, winkel2020implementation, rieger2023granular}, that improve and simplify the fabrication of superinductors. 

One of the most interesting features of fluxonium is its rich energy level structure and large anharmonicity compared to transmon, which facilitates the implementation of  fast and high-fidelity single-qubit gates with negligible leakage \cite{zhang_2021, bao_2022, nguyen_2022,somoroff_2023}. Most of the protocols tested so far to control and manipulate the fluxonium rely on the application of small amplitude and/or resonant Rabi-based   drivings \cite {zhang_2021,rower_2024}. However, in order to fully explore its energy  spectrum, an excellent alternative is the Landau-Zener-St\"uckelberg (LZS)  protocol, which  consists of driving the quantum circuit using a strong, variable-amplitude, off-resonant harmonic signal \cite{shevchenko_2010, shevchenko_2018,ivakhnenko_2023}. It was originally implemented to analyze the spectrum  of flux and charge qubits \cite{Oliver_2005,Sillanpaa_2006, Izmalkov_2008}, even beyond the two lowest energy levels \cite{Oliver_2009, Berns_2008,Ferron_2012, gustavsson_2012,Ferron_2016}. In recent years, the LZS protocol has been proposed as a tool to generate steady-state entanglement between two flux qubits \cite{Gramajo_2017,gramajo_2018, gramajo_2021}, and in cQED architectures where one \cite{Bonifacio_2020} or two flux-like qubits  \cite{gallardo_2022} are driven and coupled to a resonator in a dissipative environment. Moreover, the LZS protocol has recently been tailored to implement fast gates (single passage) on small gap qubits \cite{campbell_2020, caceres_2023} such as fluxonium, and as a strategy to improve coherence \cite{mundada_2020}. 

Despite the success of the LZS amplitude spectroscopy of flux and charge qubits, similar experiments have not been yet reported for the fluxonium. In the present work we perform numerical simulations of a LZS protocol to study the multi-level structure of the fluxonium, aiming to obtain a realistic description of its spectrum useful for experimental implementations. We analyze the LZS spectra showing the time-averaged population of the different levels as a function of the amplitude of a nonresonant drive and the magnetic flux which tunes the Hamiltonian. In particular, we focus on the resonance patterns in the spectra that can be understood as originated from constructive interferences favoring transitions to higher levels, which are unavoidably populated for large-amplitude drives. In order to emphasize the relevance of the multilevel nature of the fluxonium, we present first the numerical results of a brut-force truncation to two levels. We discuss the difference with the standard formulation of the LZS problem and derive a theoretical description of the resonance patterns using a simplified model of the avoided crossings in terms of a dynamic phase accumulation. For the multilevel case, we derive an effective two-level Hamiltonian using a Schrieffer-Wolff transformation \cite{schrieffer1966relation, bravyi2011schrieffer} starting from the Floquet Hamiltonian in the Sambe space \cite{santoro2019introduction, bukov2015universal} that reproduces relevant characteristics of the numerical spectra.

The paper is organized as follows: in Sec.\ref{sec:mh} we introduce the model Hamiltonian of a harmonically driven fluxonium. In Sec. \ref{sec:LZS_num} we present the numerical simulations of the time average population as a function of the driving amplitude and flux detuning, which we model in Sec. \ref{sec:LZS_theo}. In Sec. \ref{sec:disc} we discussion the effective model and the possibility to use LZS protocol to implement single qubit gates or to generate entanglement. In order to render the main message of this work more accessible, complementary information is given in the appendix. 

\section{Model Hamiltonian of a driven fluxonium}
\label{sec:mh}
\begin{figure}[t!]
    \centering
    \includegraphics[width=\linewidth]{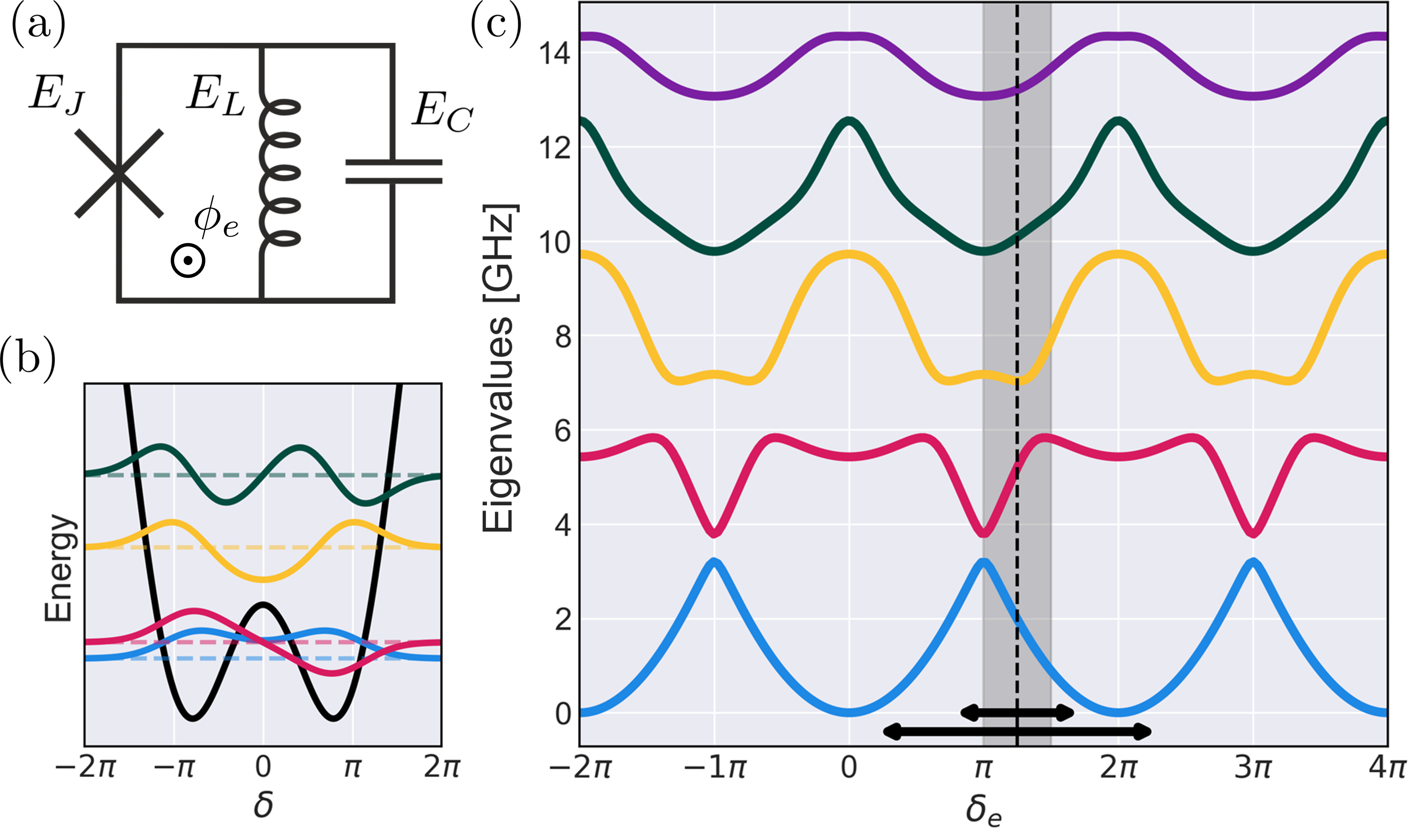}
    \caption{\textbf{Landau-Zener-Stückelberg interferometry of a fluxonium circuit:}(a) Fluxonium circuit consisting of a Josephson junction in a loop with an inductor threaded by a magnetic flux $\phi_{e}$, in parallel with a capacitor. (b) Schematic representation of the static potential energy as a function of the quantum variable $\delta$ (in black) at $\delta_{0}= \pi$ (see text) together with the first four eigenenergies (dashed lines) and the corresponding eigenstates $\psi_i(\delta)$ in arbitrary units. (c) Spectrum of the fluxonium as a function of $\delta_e$. The external flux is modulated periodically in time producing a phase $\delta_{e} \rightarrow \delta_{e}(t)=\delta_0 + A\cos(\omega t)$. The gray area represents the region where we study the LZS spectra: $\delta_0\in [\pi,1.5\pi]$, the vertical dashed line indicates an arbitrary value of phase offset $\delta_0$ and the short (long) black arrow indicates the maximum drive amplitude $A=0.4\pi$ ($A=\pi$) considered for the discussion in the main text (appendix).}
    \label{fig:fig1}
\end{figure}

The fluxonium circuit Hamiltonian is given by \cite{Manucharyan2009}
\begin{equation}
\hat{H}_0 = 4E_C \hat{n}^2 + \frac{E_L}{2} (\hat{\delta} - \delta_e)^2 - E_J \cos\hat{\delta},
\label{H_fluxonium_nodrive}
\end{equation}
where $\hat{n}$ and $\hat{\delta}$ are the conjugate quantum variables $[\hat{\delta},\hat{n}]=i$ describing the charge number and phase degrees of freedom, respectively. The relative values of the charging energy $E_C$, the inductive energy $E_L$ and the Josephson energy $E_J$ determine the operation regime of the circuit, and $\delta_e=2\pi \phi_e/\Phi_0$ is given by the external flux $\phi_e$ threading the loop formed by the Josephson junction and the inductor, in units of the flux quantum $\Phi_0$ (see Fig. \ref{fig:fig1}(a)). Following the analysis of Ref. \onlinecite{nguyen_2022}, we choose $E_J=2\pi\times4\,\text{GHz},E_L=2\pi\times1\,\text{GHz},E_C=2\pi\times1\,\text{GHz}$ which give the potential landscape shown in Fig.\ref{fig:fig1}(b), for $\delta_e=\pi$, and the energy levels displayed in Fig. \ref{fig:fig1}(c)  as a function of $\delta_e$ and represented in an ``extended zone''.

We consider an harmonic drive applied to the circuit via a fast antenna, such that $\delta_e \rightarrow\delta_e(t)=\delta_0+A\cos(\omega t)$. After replacing this modulation in Eq.(\ref{H_fluxonium_nodrive}), the driven Hamiltonian results
\begin{equation}
\hat{H}(t) = 4E_C \hat{n}^2 + \frac{E_L}{2} (\hat{\delta} - \delta_0)^2 - E_J \cos\hat{\delta} - E_L \hat{\delta} A \cos(\omega t).
\label{H_fluxonium}
\end{equation}

For each $\delta_0$, we diagonalize numerically the Hamiltonian $\hat{H}_0$ and project the driven Hamiltonian of Eq.(\ref{H_fluxonium}) in the resulting eigenbasis getting
\begin{equation}
\hat{H}(t) = \sum_i E_i |i\rangle\langle i | + \sum_{ij} A E_L \delta_{ij}  |i\rangle \langle j| \cos(\omega t),
\label{H_projected}
\end{equation}
where we simplify the notation defining $\{|i\rangle\}\equiv\{|i(\delta_0)\rangle\}$, the eigenenergies $E_i\equiv E_i(\delta_0)$ and the matrix elements $\delta_{ij}\equiv\langle i(\delta_0)|\hat{\delta}|j(\delta_0) \rangle$.

From the  Hamiltonian  Eq.(\ref{H_projected})  we integrate numerically the time evolution for a fixed frequency $\omega$ and amplitude $A$ of the drive, keeping the lowest $N_{\text{levels}}$, and compute the instantaneous population of each state $P_i(t,A,\delta_0)$. We define the (time) averaged population of each level after $N_p$ periods of the drive as
\begin{equation}
\overline{P}_i(A,\delta_0) = \frac{1}{T}\int_0^{T} P_i(t,A,\delta_0) dt,
\end{equation}
being  $T=N_p \tau$ with  $\tau= 2\pi/{\omega}$ the driving period. 

\section{LZS spectra: numerical results}
\label{sec:LZS_num}

\begin{figure}[t!]
\centering
\includegraphics[width=1\columnwidth]{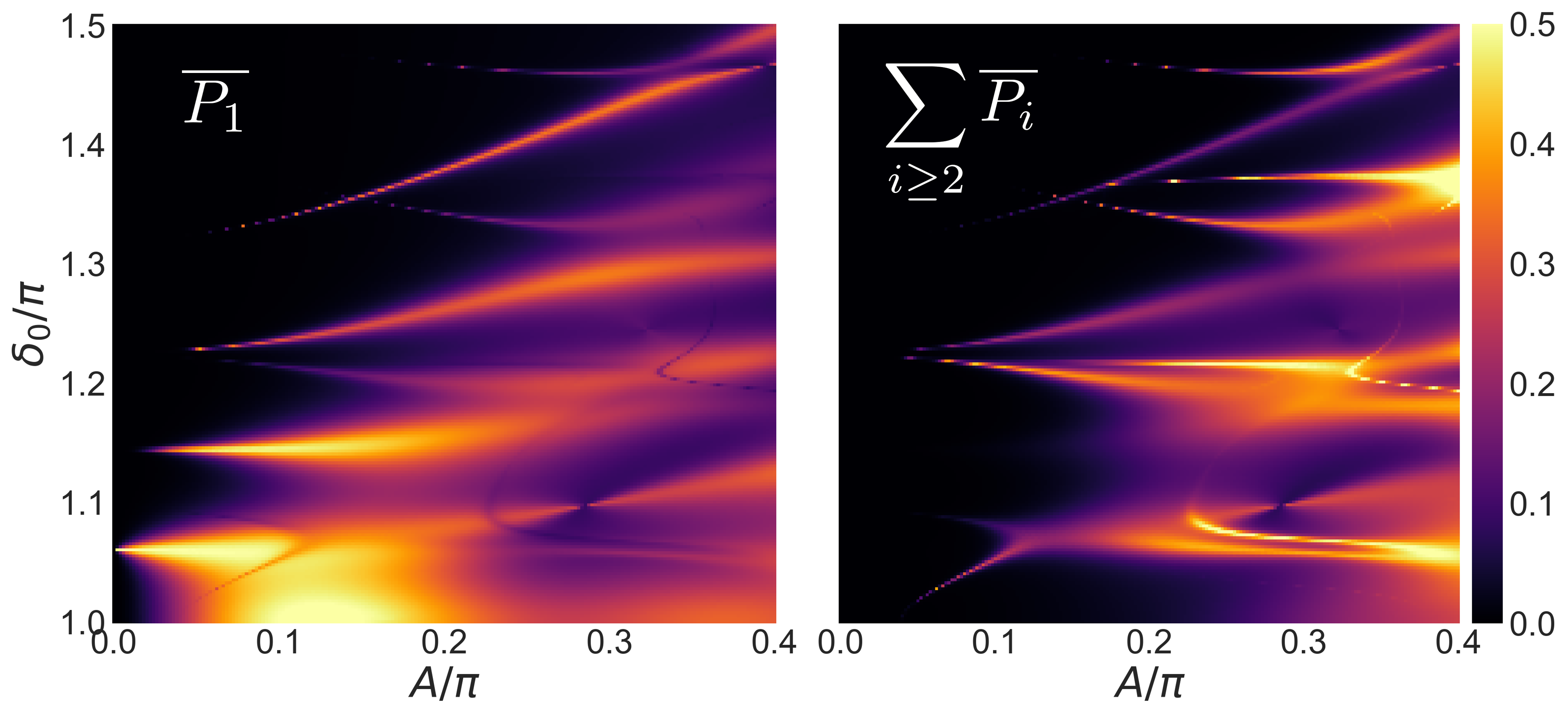}
    \caption{\textbf{LZS spectrum:} Time-averaged population of the driven fluxonium as a function of drive amplitude $A$ and the external flux $\delta_0$. Averaged population of (left) the first excited state $|i=1\rangle$ and (right) the non-computational subspace $\sum_{i\ge 2} \overline{P}_i(A,\delta_0)$. The dynamics were simulated using a Hamiltonian truncated to $N_{\text{levels}} = 10$ and averaged over 500 drive periods, always taking the ground state as the initial state. The chosen drive frequency is $\omega=2\pi \times 1\,\text{GHz} \approx 1.7 \omega_{01}(\pi)$}
    \label{fig:fig2}
\end{figure}

In Fig. \ref{fig:fig2} we present the numerical results for the LZS spectra (i.e. the averaged populations as a function of $A$ and $\delta_0$) of the first excited state $\overline{P}_1(A,\delta_0)$ in (a) and the sum for the states out of the computational subspace $\sum_{i\ge 2}\overline{P}_i(A,\delta_0)$ in (b) with $A\in [0,0.4\pi]$ and $\delta_0\in[\pi,1.5\pi]$. The fluxonium is driven with a frequency of $\omega=2\pi \times 1\,\text{GHz}$ which is $\approx 1.7 \omega_q$, where as usual the qubit frequency is defined as $\hbar\omega_q=E_1(\pi)-E_0(\pi)$ at the sweet spot. The numerical integration of the Schrödinger equation was performed using the Quantum Toolbox in Python (QuTiP) package \cite{johansson2012qutip}. We have verified that, within this regime, the transition probabilities can be accurately determined by numerically evolving the system -initially in the ground state- using a Hamiltonian truncated to $N_{\text{levels}} = 10$ (see App.\ref{ap:LZS_full_spectra}).  We focus here on moderate drive amplitudes $A\lesssim 0.4\pi$, a larger amplitude range is presented in App. \ref{ap:LZS_full_spectra}. We have verified that including relaxation does not modify significantly the results since the integration time is $T=500\ $ns, much smaller than typical $T_1\sim$300\ $\mu$s in fluxonium \cite{zhang_2021, somoroff_2023}. 

Figure \ref{fig:fig2} shows the presence of ``resonances'', defined as maxima of the time average populations, in analogy to resonant 2-level transitions, occurring at values of $\delta_0$ such $E_1(\delta_0)-E_0(\delta_0)\sim n\hbar\omega$, which start to depart from this condition as $A$ increases. There are also features that correspond to resonances with the level $|2\rangle$ (or higher, as seen in the right panel). These LZS-resonances are originated from a constructive interference that maximizes the occupation probability of an excited state. The patterns exhibit clear deviations from those predicted for a two level system (TLS) where a single avoided crossing with a gap $\tilde{\Delta}$ is considered \cite{shevchenko_2010} (see App.\ref{ap:LZS_classic}). When a TLS is driven by a periodic signal of amplitude $A$ and frequency $\omega$ through an avoided crossing, the resonance condition for which the transfer of population is maximum depends on the velocity of passing. Qualitatively, the slow driving regime is defined for $A \omega < \tilde{\Delta}^2$, while the fast driving condition is attained for $A \omega \gg \tilde{\Delta}^2$. In this last case, for which the Landau-Zener probability is large, the theory predicts resonant (straight) lines in the LZS patterns appearing when the energy separation between both levels matches that of an integer number of drive photons, modulated in amplitude by Bessel functions of the first kind. On the other hand, in the slow driving regime the LZS resonances define arcs, but explicit analytical expressions for the resonance condition are lacking \cite{shevchenko_2018}. These LZS resonance patterns have been studied and probed in driven qubits restricted to TLS in both regimes \cite{stehlik_2012, koski_2018, gramajo_2020}.

In order to get an understanding of the features observed in the LZS patterns of the fluxonium, we develop in the next two sections a theoretical framework based on effective models. We first simulate the time-evolution with the Hamiltonian brut-force truncated to $N_{\text{levels}}=$2 to highlight the differences with the TLS standard LZS analysis. As we show, this oversimplification is not enough to account for the spectrum depicted in  the left panel of Fig. \ref{fig:fig2}, although  some  features of the lower resonances are well captured. Next, we construct an effective two-level system Hamiltonian using a Schrieffer-Wolf transformation that keeps to some extent the effect of the multi-level structure. We identify the parameter regions where a large non-resonant amplitude drive can be used to implement fast LZS single qubit gates and also entanglement protocols in a realistic situation.

\begin{figure}[t!]
    \centering
\includegraphics[width=0.9\columnwidth]{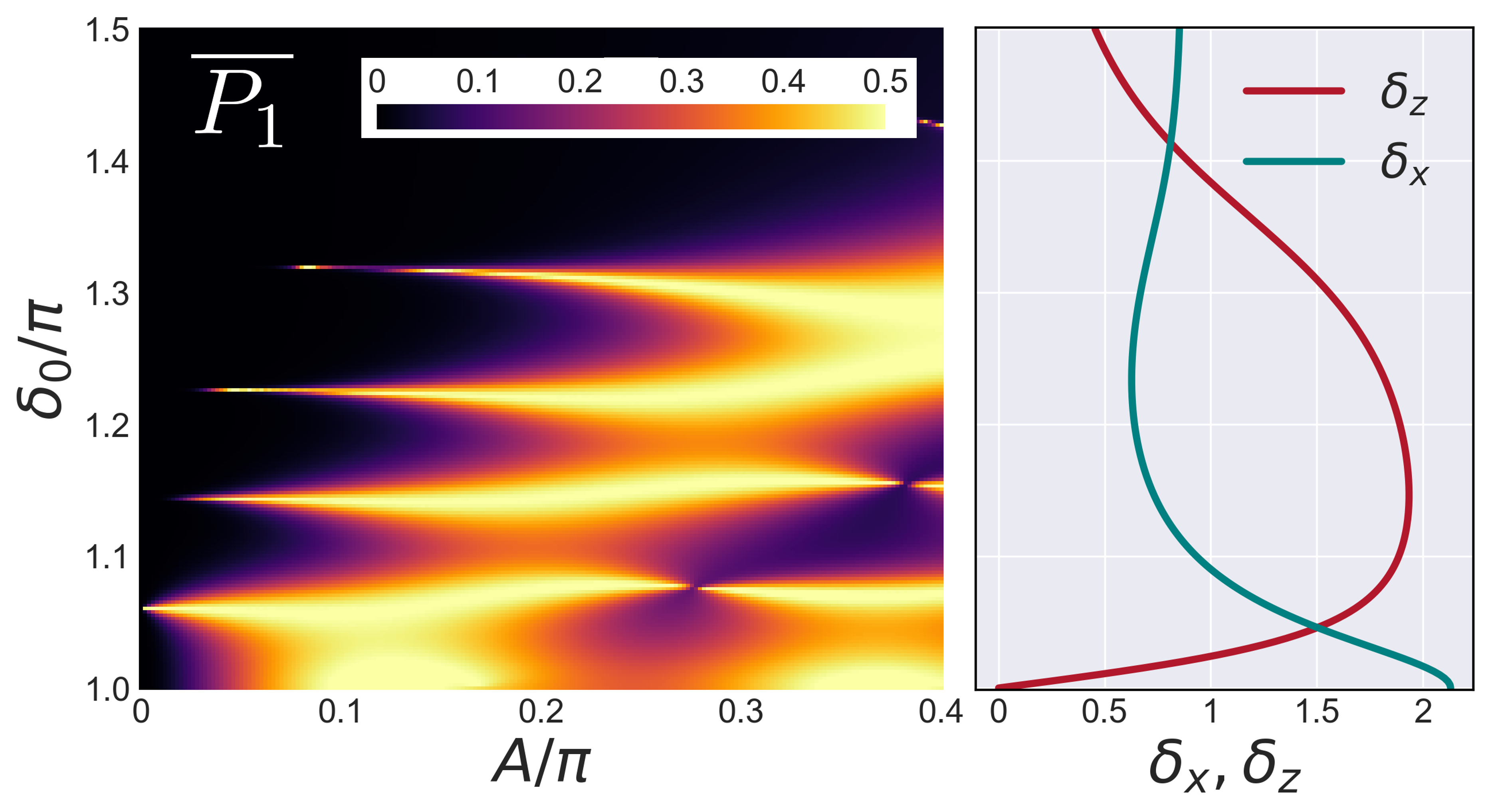}
    \caption{\textbf{LZS spectra for the brut-force truncated two-level fluxonium} Left: Time-averaged population of the excited state $\overline{P}_1$ as a function of the drive amplitude $A$ and the phase offset $\delta_0$. Right: Matrix elements $\delta_x$ and $\delta_z$ as defined in the text, as a function of $\delta_0$.}
    \label{fig:fig3}
\end{figure}

\section{LZS spectra: theoretical modeling}
\label{sec:LZS_theo}
\subsection{2-level Hamiltonian from brut-force truncation}
We start by analyzing the time evolution of the fluxonium in a qubit-like  approximation, i.e. keeping only $N_{\text{levels}}=$2 in Eq.(\ref{H_projected}). In the qubit eigenbasis $\{|0(\delta_0)\rangle,|1(\delta_0)\rangle\}$, Eq. \eqref{H_fluxonium} can be expressed as,
\begin{equation}
    \hat{H}_{\text{TL}}(t) = \frac{\omega_q(\delta_0)}2 \sigma_z - E_L A [ \sigma_x \delta_x(\delta_0) + \sigma_z\delta_z(\delta_0) ] \cos(\omega t),
    \label{H_complicado}
\end{equation}
where $\delta_x(\delta_0) = \langle 0(\delta_0) | \hat{\delta} | 1(\delta_0)\rangle$, $\delta_z(\delta_0) = \frac 12 (\langle 0(\delta_0) | \hat{\delta} | 0(\delta_0)\rangle - \langle 1(\delta_0) | \hat{\delta} | 1(\delta_0)\rangle)$, making explicit the dependence on the operation point $\delta_0$. The matrix elements as a function of $\delta_0$ are shown in the right panel of Fig. \ref{fig:fig3}. 

The LZS spectrum $\overline{P_1}(A,\delta_0$) obtained from the numerical integration of the Schrödinger equation after the brut-force two-level truncation is shown in the left panel of Fig.\ref{fig:fig3}. It can be compared with the result shown in Fig.\ref{fig:lzs(t)}(b) for the simple situation of a TLS with an anticrossing, as  described in App.\ref{ap:LZS_classic}. Using the energies $E_i(\delta_0)$ in Fig.\ref{fig:fig1}(c), we have verified that the resonances correspond approximately to $E_1(\delta_0)-E_0(\delta_0)=n\hbar\omega$, but there are some differences: on the one side, the positions of the resonances depend on the drive amplitude, and second, they disappear as $\delta_0\gtrsim 1.4\pi$. The key to understand these features is to consider the functional form of the drive in the Hamiltonian of Eq.(\ref{H_complicado}), and the dependence on $\delta_0$ of the matrix elements depicted in the right panel of Fig. \ref{fig:fig3}. It can be observed that the drive acts both on $\sigma_x$ and $\sigma_z$, which does not correspond to a typical Rabi-like or LZS-like scenario. 

For  given $(\delta_0,A)$ and defining the rotation angle $\theta = \arctan{({\delta_x(\delta_0) / \delta_z(\delta_0))}}$, we  build an effective LZS-like Hamiltonian, which  in the rotated basis results
\begin{eqnarray}
    \hat{H}'_{\text{TL}}(t) &=& \frac12\hbar\omega_q(\delta_0) (\cos(\theta) \sigma'_z + \sin(\theta) \sigma'_x) \nonumber\\&+& E_L A \sqrt{\delta_x^2(\delta_0) + \delta_z^2(\delta_0)} \cos(\omega t) \sigma'_z.
    \label{H_lzz_rot}
\end{eqnarray}
This is the Hamiltonian of a longitudinally driven TLS, as the one considered in App. \ref{ap:LZS_classic}  \cite{shevchenko_2010,shevchenko_2018}, but characterized by the effective parameters 
\begin{eqnarray*}
\Delta_{\text{eff}} &=& \hbar\omega_q \sin(\theta), \quad 
\varepsilon_0 = \hbar\omega_q \cos(\theta) \\
A_{\text{eff}} &=& 2A E_L\sqrt{\delta_x^2(\delta_0) + \delta_z^2(\delta_0)},    
\end{eqnarray*}
which depend on $\delta_0$ and $A$. In the fast driving regime, the  LZS spectrum  can be obtained using the expression
\begin{equation*}
    \overline{P}_{1}(\varepsilon_0,A_{\text{eff}},\omega) = \frac 12  \sum_{k = -\infty}^\infty \frac{\left(\Delta_{\text{eff}} J_k\left( \frac {A_{\text{eff}}}{\hbar\omega} \right)\right)^2}{(\varepsilon_0 - k\hbar\omega)^2 + \left(\Delta_{\text{eff}} J_k\left( \frac {A_{\text{eff}}}{\hbar\omega} \right)\right)^2},
\end{equation*}
derived from the Adiabatic Impulse Model (AIM) in App. \ref{ap:LZS_classic}. As shown in Fig. \ref{fig:LZS_2ls}, the effective LZS-like model is in very good agreement with the numerical result. The AIM fails to describe the intermediate driving velocity regime, in particular when the amplitude is not enough to arrive to the anti-crossing. 

The disappearance of the LZS resonances for $\delta_0 \gtrsim 1.4\pi$ is primarily due to the dependence of the matrix elements $\delta_x(\delta_0)$ and $\delta_z(\delta_0)$ on $\delta_0$, which alters the angle $\theta$. As seen in the spectrum of Fig. \ref{fig:fig1}c, beyond the shoulder at $\delta_0 \approx 1.4\pi$, the driving becomes predominantly orthogonal to the time-independent component of the Hamiltonian, leaving only a strongly detuned, Rabi-like contribution.

In order to better understand the spectra presented in Fig. \ref{fig:fig2}, we initially attempted a progressive truncation approach, starting with two levels, then three and so on (see App. \ref{ap:LZS_2ls} \footnote{At a first glance, the two-level truncation seems to be a quite reasonable approach at the operationally ``sweet spot'', $\delta_0=\pi$, since $E_2(\pi)-E_1(\pi) \gg E_1(\pi)-E_0(\pi)$. For the sake of comparison, the qubit Hamiltonian at $\delta_0=\pi$ is $\hat{H}\approx \frac{\omega_q(\pi)}2 \sigma_z - 2 E_L A \sigma_x \cos(\omega t)$, such that the Rabi frequency for a resonant tone ($\omega=\omega_q(\pi)$) is $\Omega_R\approx 4 E_L A$. This means that the amplitude to realize a PI-pulse in 10\ ns would be $A\approx 0.004 \pi$ ($\Omega_R=2\pi \times 0.05\ $GHz). As we are considering non-resonant tones and much larger amplitudes, the full LZS spectrum (see Fig. \ref{fig:fig3}) differs significantly from what is shown in Fig. \ref{fig:fig2}}). However, the resulting spectra proved difficult to reconcile with simplified models of superimposed transitions arising from AIM, considering pairs of levels such as $|0\rangle \leftrightarrow |1\rangle$, $|0\rangle \leftrightarrow |2\rangle$, and $|1\rangle \leftrightarrow |2\rangle$. As we show next, we developed an effective two-level model that incorporates the influence of higher-energy states and accurately describes the observed spectra.

\subsection{Effective low-energy two-level Hamiltonian for the driven fluxonium}
In this section we use a method referred to as the Floquet-Schrieffer-Wolff (FSW) method \cite{santoro2019introduction, bukov2015universal}, to obtain an effective two-level Floquet Hamiltonian that incorporates the influence of higher energy levels in the limit of low driving amplitudes.

Since the Hamiltonian Eq.(\ref{H_projected}) is periodic in time (i.e. $\hat{H}(t)=\hat{H}(t+\tau)$), we employ the Floquet formalism \cite{shirley_1965,kohler_1997,son_2009,Ferron_2012,Ferron_2016}, where the solutions of the time dependent Schr\"odinger equation are expressed as $\ket{\psi_\alpha(t)}=e^{-i\frac{\varepsilon_\alpha t}{\hbar}}\ket{\alpha(t)}$. Here, the Floquet states $\ket{\alpha(t)}$, and the corresponding quasienergies, $\varepsilon_\alpha$, are obtained from the eigenvalue equation $\hat{H}_F\ket{\alpha(t)}= \varepsilon_\alpha\ket{\alpha(t)}$, being $\hat{H}_F\equiv \hat{H}(t)-i\hbar\partial_t$ the Floquet Hamiltonian. As the Floquet states satisfy $\ket{\alpha(t)}=\ket{\alpha(t+\tau)}$ they can be Fourier decomposed \cite{son_2009}. Therefore, in the extended Hilbert or Sambe space, which is the tensor product of the Hilbert space of the undriven system and the space of time periodic functions  \cite{grifoni_1998}, an  equivalent time independent but infinitely dimensional Floquet matrix eigenvalue problem is obtained \cite{son_2009} (see details in App. \ref{ap:FSW_app}).

\begin{figure*}[t!]
    \centering
\includegraphics[width=\textwidth]{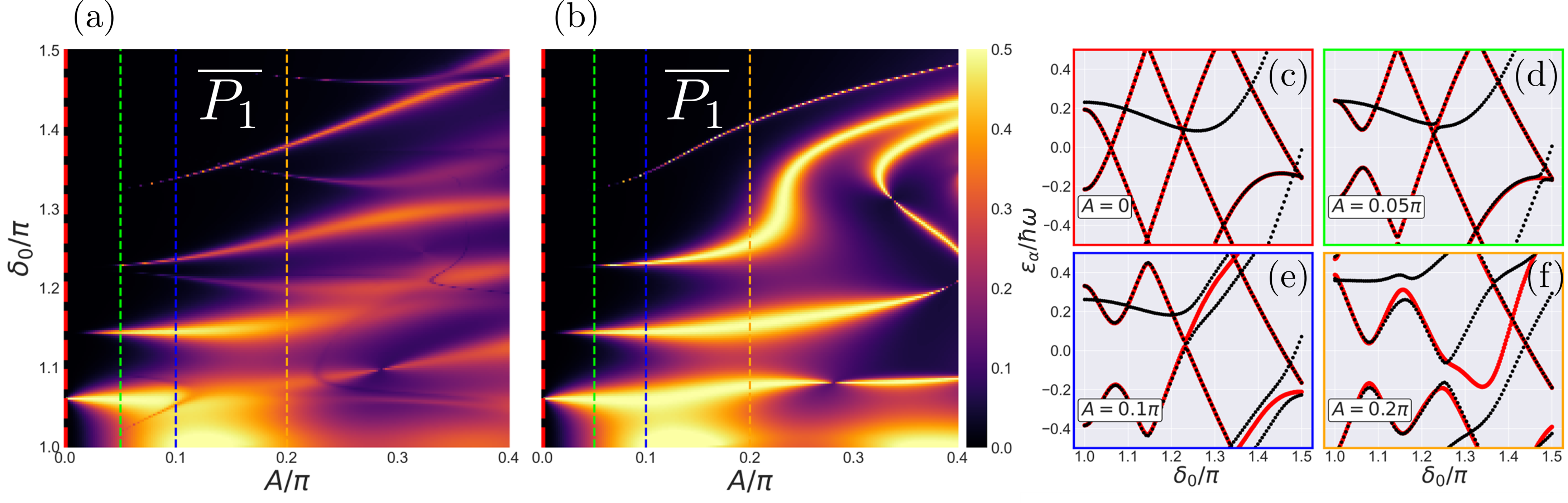}
    \caption{\textbf{Effective model for the fluxonium LZS spectra:} (a),(b): Time averaged population of the first excited state $\overline{P}_1$ as a function of $\delta_0$ and the driving amplitude $A$.  (a) full simulation and (b) effective TLS obtained from FSW method (see text). (c)-(f) Quasienergies restricted to the first Brillouin zone as a function of $\delta_0$ calculated for the driven fluxonium (black dots) for the values of $A$ indicated with dashed color lines in (a) and (b). For clarity, we show only the quasienergies that for $A=0$ can be associated to the lowest 3 levels of the fluxonium. With red lines we show the result of the eigenenergies of the effective two-level FSW Hamiltonian. In this range of amplitudes, the largest contribution to the FSW correction is given by the second excited state $|2\rangle$, the effect of higher levels is much smaller, and resulted in identical spectra as in (b).}
    \label{fig:fig4}
\end{figure*}

For the present case of a periodic drive $\sim \hat{\delta} \cos(\omega t)$, the time independent Floquet Hamiltonian written in the extended Sambe basis $|i,N\rangle\equiv|i\rangle \otimes |N\rangle$, where $N$ represents the Fourier or ``photon'' index, is given by 
\begin{eqnarray*}
\hat{H}_F &=& \sum_{i,N} \left(E_i+N\hbar \omega\right) |i,N\rangle\langle i,N | \nonumber \\&+& \sum_{ij,N}  \langle i |\tilde{\delta}|j\rangle \left(|i,N+1\rangle \langle j,N| + |i,N-1\rangle \langle j,N| \right),
\label{Hfloquet}
\end{eqnarray*}
where $\tilde{\delta}_{ij} \equiv ( A E_L/2 )\hat{\delta}_{ij} $. Here we observe that the fluxonium levels are mixed exclusively by the drive.

In order to build the effective two-level Hamiltonian, we introduce the projectors $\hat{P}_{q} = (|0\rangle\langle0| + |1\rangle\langle 1|) \otimes \mathds 1_{N_F}$, where $N_F$ is the dimension of the "photon" space (in principle, infinite), and $\hat{P}_{\overline{q}} = \mathds 1_{N_{\text{levels}}} \otimes {\mathds 1}_{N_F} - \hat{P}_q$, projecting onto the qubit subspace $\{|0N\rangle,|1N\rangle\}$ and on the remaining $\{|jN\rangle\}$ states with $j\ge 2$, respectively. The Floquet Hamiltonian of the full Hilbert space can be decomposed (see App. \ref{ap:FSW_app}) as $\hat{H}_F= \hat{H}_q + \hat{H}_{\overline{q}} + \hat{V}$, where
\begin{eqnarray*}
\hat{H}_q &=& \sum_{i=0,1;N} \left(E_i+N\hbar \omega\right) |i,N\rangle\langle i,N | \nonumber \\&+& \sum_{i,j=0,1;N}  \langle i |\tilde{\delta}|j\rangle \left(|i,N+1\rangle \langle j,N| + |i,N-1\rangle \langle j,N| \right),
\end{eqnarray*}
and where $\hat{V}$ includes all matrix elements that couple states $\{|0N\rangle,|1N\rangle\} $ and $\{|jN\rangle\} $, with $j\ge2$, thereby linking the subspaces $q$ and $\overline{q}$,
\begin{eqnarray*}
\hat{V} &=& \sum_{\substack{i=0,1\\j\ge2,N}} \langle i |\tilde{\delta}|j\rangle \left(|i,N+1\rangle \langle j,N| + |i,N-1\rangle \langle j,N| \right) + h.c.
\end{eqnarray*}

The elements of $\hat{V}$ are linear in the amplitude $A$ and non-diagonal, involving transitions in the photon blocks $N\rightarrow N\pm 1$. To get an effective TLS Hamiltonian we would like to treat $\hat{V}$ as a perturbation and decouple from the high-energy levels. We can do this using a Schrieffer-Wolff transformation with some subtleties. First we present the result and then discuss its validity. To second order in $\hat{V}$, and assuming a block-diagonal unperturbed Hamiltonian, we get (see App.\ref{ap:FSW_app})
\begin{equation*}
\tilde{H}_q^N = \left(\begin{array}{cc}
     \tilde{E}_0 \pm N\hbar \omega& \tilde{E}_{01} \\
     \tilde{E}_{01} & \tilde{E}_1\pm N\hbar
\end{array}\right),
\end{equation*}
where
\begin{eqnarray*}
\tilde{E}_{i}&=&E_{i}+\sum_{j\ge2 }\tilde{\delta}_{ij}\tilde{\delta}_{ji}\left(\frac{1}{E_i-E_j-\hbar\omega}+\frac{1}{E_i-E_j+\hbar\omega}\right),  
\end{eqnarray*}
with $i=0,1$ and 
\begin{eqnarray*}
\tilde{E}_{01}&=&\frac{1}{2}\sum_{j\ge 2 }\tilde{\delta}_{0j}\tilde{\delta}_{j1}\left(\frac{1}{E_0-E_j-\hbar\omega}+\frac{1}{E_0-E_j+\hbar\omega} \right.\\
&+&\left.\frac{1}{E_1-E_j-\hbar\omega}+\frac{1}{E_j-E_i+\hbar\omega}\right).  
\end{eqnarray*}

After the SW transformation there are no corrections to the terms
\begin{eqnarray*}
\sum_{i,j=0,1,N}  \tilde{\delta}_{ij} \left(|i,N+1\rangle \langle j,N| + |i,N-1\rangle \langle j,N| \right),
\end{eqnarray*}
therefore it is straightforward to obtain the time dependent Hamiltonian 
\begin{equation}
    \hat{H}_{\text{FSW}}(t) = \tilde{H}_q + E_L A \hat{\delta} \cos(\omega t),
\end{equation}
with
\begin{equation*}
\tilde{H}_q = \left(\begin{array}{cc}
     \tilde{E}_0 & \tilde{E}_{01} \\
     \tilde{E}_{01} & \tilde{E}_1
\end{array}\right).
\end{equation*} 

The LZS spectra calculated by the numerical integration of this effective model is presented in Fig.\ref{fig:fig4}(b) and compared to the one obtained from the full numerical simulation in Fig.\ref{fig:fig4}(a). Noticeable, the positions of the resonances and their displacements as a function of the amplitude $A$ are reproduced for values A$\lesssim 0.3 \pi$, as well as the intensities. In Fig.\ref{fig:fig4}(c)-(f), we compare the quasienergies calculated for the driven fluxonium (black dots) with the eigenenergies of the effective TLS FSW Hamiltonian in the Sambe space (red lines), which under this approach, play the same role as the quasienergies. The panels correspond to the values of $A$ indicated with dashed color lines in (a) and (b). For clarity, we show only the quasienergies that for $A=0$ can be associated to the lowest 3 levels of the fluxonium. The effective model reproduces very well the correction of the two-level resonance patterns produced by the virtual transitions to high-energy states. It starts to fail for large amplitudes. Furthermore, the patterns due to resonances  with higher energy states are not accounted for by the effective TLS FSW Hamiltonian (See App. \ref{ap:LZS_2ls} for more details).

Let us now discuss the subtleties. First, the SW transformation assumes that the subspaces $q$ and $\overline{q}$ are well-separated in energy. Since we are working in the Sambe space, there are crossings of energies where the transformation is not valid. For example, the energy of the level $j=2$ translated into the first Brillouin zone by $N\hbar \omega$ can cross the energies of the levels in the two-level subspace $i=0,1$ (see Fig. \ref{fig:fig4}(c)). Second, to obtain the effective Hamiltonian of the  form   $\tilde{H}_q$, we need the unperturbed Hamiltonian  to be  block-diagonal. More precisely the expressions obtained for $\tilde{E}_{i}$ and $\tilde{E}_{01}$,
consider  $E_i+N\hbar\omega$ to be the energy of the state $|iN\rangle$ for $\hat{V}=0$. This is correct as the perturbation is zero for $A=0$, but for finite values of $A$, the Hamiltonian $\tilde{H}_q$ is not diagonal. Third, in the second-order approximation in $\hat{V}$, terms appear that couple blocks differing by two photons, that is, transitions such that $|0 N\rangle \rightarrow |2 N\pm 1\rangle \rightarrow |0 N\pm 2\rangle$. These contributions give rise to a term $\hat{J}$ in the Hamiltonian proportional to $A^2$ that will contribute to $\exp(\pm i2\omega t)$ in the corrected time-dependent form. We have verified that these contributions remain negligible in the range of amplitudes where the approximation holds.

\begin{figure}[t!]
\centering
\includegraphics[width=0.9\columnwidth]{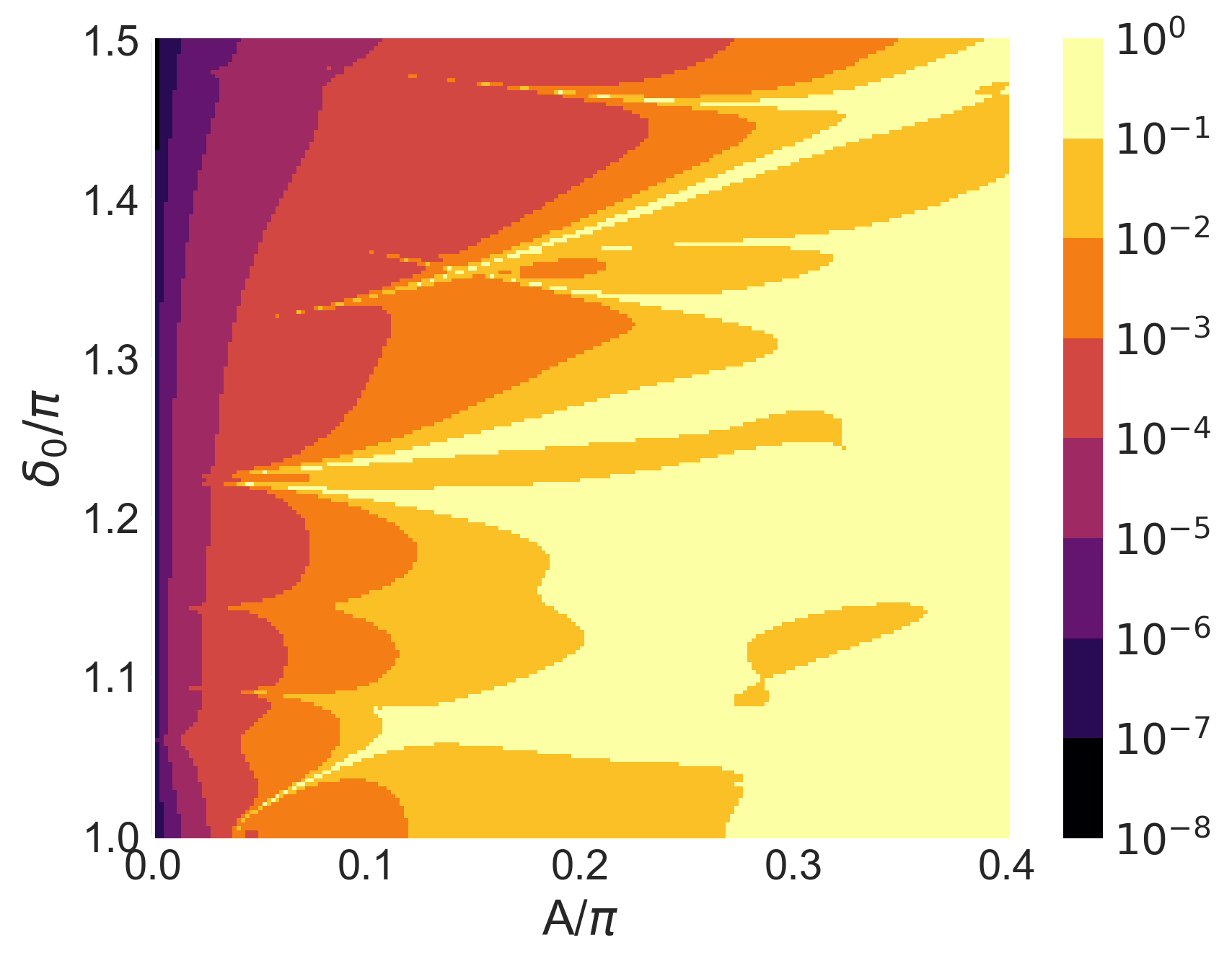}
    \caption{\textbf{Leakage:} Total time-averaged population out of the computation subspace $\sum_{j\ge 2} \overline{P}_j$ as a function of driving amplitude $A$ and phase detuning $\delta_0$ in the same range of Fig. \ref{fig:fig4}(a). Regions exhibiting leakage to higher energy levels $\lesssim 10^{-3}$ are observed, indicating the potential for high-fidelity quantum gate design.}
    \label{fig:fig5}
\end{figure}

\section{Discussion and Conclusions}
\label{sec:disc}
In this work we have studied the Landau-Zener-Stückelberg interferometry of a fluxonium circuit, fully accounting for the multi-level structure of its Hamiltonian. Our main goal was to provide realistic simulations of the expected LZS spectra. These spectra are indeed rich and complex, and for moderate driving amplitudes, we derived an effective two-level system (TLS) model using a Floquet-Schrieffer-Wolff approach. This effective model successfully captures the critical role of high-energy states, and we identify regions in the parameter space of Figs. \ref{fig:fig4}(a,b) where it offers a reliable description. In particular, as shown in Fig. \ref{fig:fig5} the total time-averaged population outside the computational subspace remains below 10$^{-3}$ for driving amplitudes $A<0.1 \pi$. Within these regions fast single qubit gates could be feasible \cite{campbell_2020, caceres_2023} as well as entanglement protocols based on a LZS approach \cite{gramajo_2021, gallardo_2022}. This would be suitable in particular for small-gap qubits \cite{campbell_2020}. For a realistic determination of gates fidelities, dissipation due to coupling with the environment should be also considered. Interestingly, such dissipation could partially suppress leakage and, in turn, enhance gate performance. Overall, our results lay the groundwork for leveraging LZS protocols in fluxonium-based architectures, bringing us closer to faster and more robust quantum operations.

\begin{acknowledgments}
Early steps of this study where done during the master thesis of Lucas Brugevin at Instituto Balseiro. We acknowledge support from CNEA, CONICET, ANPCyT (PICT 2019-0654), UNCuyo (06/C026-T1), PIP 11220200101825CO TOSI.
\end{acknowledgments}

\appendix
\counterwithin{figure}{section}
\section{LZS for two level anticrossing}
\label{ap:LZS_classic}
Landau-Zener-Stückelberg transitions describe non-adiabatic population transfer between two energy levels when a system parameter is swept through an anticrossing in the spectrum. In the following, we highlight some important results which are useful along this work. For a review on the subject see Refs. \onlinecite{shevchenko_2010,shevchenko_2018,ivakhnenko_2023}.

\begin{figure*}[t!]
\centering
\includegraphics[width=1.8\columnwidth]{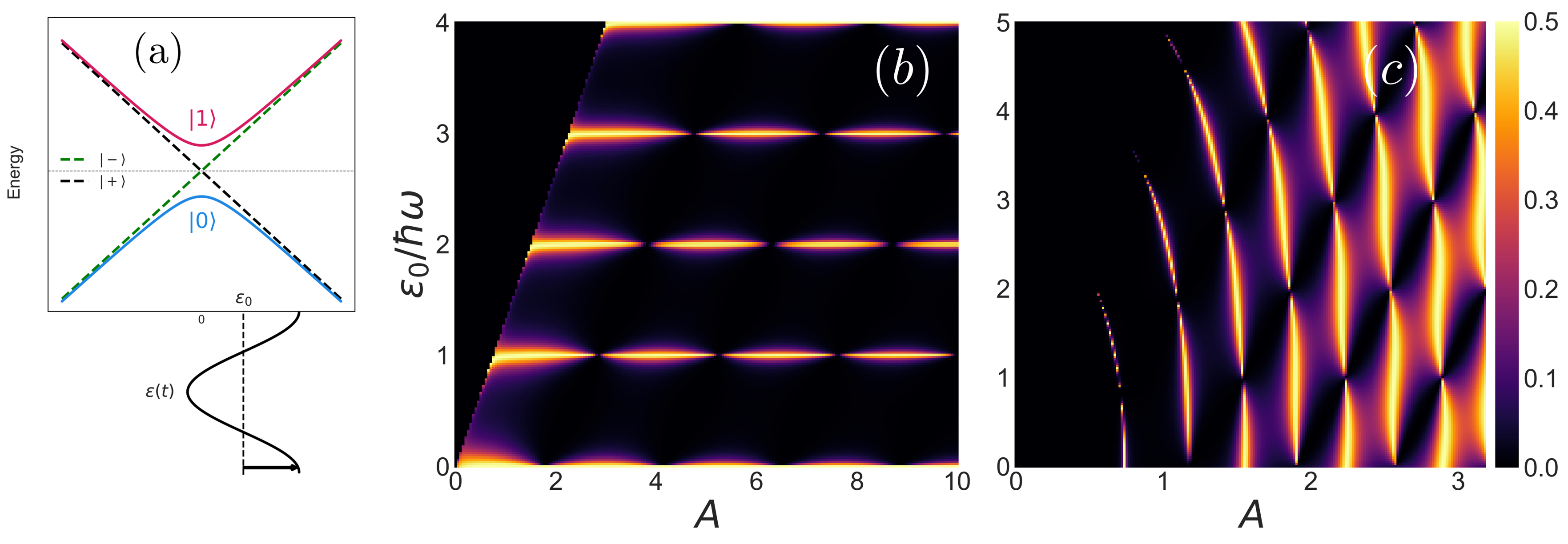}
    \caption{\textbf{2-level system LZS spectra:} (a) Schematic representation of a driven TLS: the energies are shown as a function of the detuning parameter $\varepsilon$ with an anticrossing at $\varepsilon=0$ opening a gap $\tilde{\Delta}$. The eigenstates of the uncoupled base are $|\pm\rangle$ with energies indicated with dashed lines. In the bottom the driving is produced by a periodic modulation $\varepsilon(t)$, centered at the offset $\varepsilon_0$ with amplitude $A$. (b-c) LZS spectra: Time-averaged population of $|1\rangle$ as a function $A$ and $\varepsilon_0$. In (b) $ A\omega/\Delta^2 \sim 40$ and in (c) $ A\omega/\Delta^2 \sim 0.05$, corresponding to the fast and slow driving cases, respectively.}
    \label{fig:lzs(t)}
\end{figure*}

As shown in Fig. \ref{fig:lzs(t)}(a) the ``standard'' LZS formulation of the problem considers a simple TLS with a Hamiltonian
\begin{equation}
    \hat{H}_{LZS}(t) = - \frac{\varepsilon(t)}{2} \sigma_z + \frac{\Delta}2 \sigma_x,
    \label{h_lzs(t)}
\end{equation}
where $\{|0\rangle,|1\rangle\}$ is the \textit{adiabatic} basis, eigenstates of the instantaneous Hamiltonian and $\{|-\rangle,|+ \rangle\}$ is the \textit{diabatic} basis, corresponding to the eigenstates for $\Delta = 0$.  These basis asymptotically coincide for $|\varepsilon|\gg 0$, far from the position of the  gap $\Delta$. For adiabatic processes, when $\varepsilon(t)$ varies very slowly, the system evolves following the instantaneous eigenstates. For moderate velocities $v=\dot{\varepsilon}$, of passage through the anti-crossing, there is a non-zero probability of having a Landau-Zener transition to the excited state
\begin{equation}
    P_{LZ} = \exp\left(-\frac{2}{\hbar}{\rm Im} \int_0^{t_0} (E_1(t') - E_0(t')) dt'\right),
\end{equation}
where $t_0$ is the solution to the equation $E_0(t_0) = E_1(t_0)$. Here,
\begin{equation}
    E_{0,1}(t) = \pm \sqrt{(vt)^2 + \Delta^2},
\end{equation}
taking $\varepsilon(t)=vt$. We obtain $t_0 = i\frac{\Delta}{v}$, thus
\begin{equation}
    \int_0^{i\frac{\Delta}{v}} \sqrt{(vt')^2+\Delta^2} dt' = i \frac{\pi\Delta^2}{8 v},
\end{equation}
and 
\begin{equation}
    P_{LZ} = e^{-2\pi \delta},\quad \delta = \frac{\Delta^2}{4\hbar v}.
\end{equation}

We recover the intuitive Landau-Zener criterion, when $ \hbar v \ll \Delta^2$, we have the adiabatic regime and $P_{LZ} \approx 0$. When $ \hbar v \gg \Delta^2$, it is the fast driving regime, $P_{LZ}$ is high and there are transitions to the excited state. Semi-classically, the system remain in the diabatic state, if it stated in $|-\rangle$, it will just remain in this state, which can be seen as a transition $|0\rangle\rightarrow |1\rangle$. In an intermediate velocity, $ \hbar v \le\Delta^2$, the driving regime is slow, but not adiabatic, there can be transitions to the excited state although $P_{LZ} $ is small.
As a first step to treat the periodic drive, we have to take into account coherence. After a single passage, the state is in a quantum superposition,
\begin{equation}
    |\psi(\infty)\rangle = \sqrt{1-P_{LZ}}|0\rangle + e^{i\phi_S}\sqrt{P_{LZ}} |1\rangle
\end{equation}
where $\phi_S$ is the Stokes phase, acquired after passing through the gap (this can be seen as an beam splitter \cite{shevchenko_2010})
\begin{equation}
    \phi_S(\delta) = \frac{\pi}4 + \delta(\ln(\delta) - 1) + \text{arg}\left( 
\Gamma(1 - i\delta) \right).
\end{equation}
This phase has no consequence in the populations except when the crossing happens many times. We are interested in a periodic modulation $\varepsilon(t) =  \varepsilon_0+A\cos(\omega t) $, where $\hbar\omega \neq \Delta$ and $A$ might be large. Let us  consider that at $t_0$, the system is $|0\rangle$ and $\varepsilon(t_0) = \varepsilon_0 + A$. Taking into account the evolution during a period $T = \frac{2\pi}{\omega}$, it is possible to obtain an analytic expression for the probability of being in the state $|1\rangle$ at $t_1=T$ using the Adiabatic Impulse Model (AIM)\cite{shevchenko_2010, ivakhnenko_2023}. AIM assumes that population changes occur only near the anticrossing (exactly at $\varepsilon(t) = 0$, and elsewhere the system only accumulates a dynamic phase. In a double passage, the evolution can be decomposed simply as $t_0 \rightarrow t_2$, dynamic phase accumulation
\begin{equation}
    \zeta_1 =\frac{1}{\hbar}\int_{t_3}^{t_2 + T} \sqrt{\varepsilon(t')^2 + \Delta^2} dt',
\end{equation}
in $t_2$ the anti-crossing is reached, $t_2 = \frac 1\omega \arccos(-\varepsilon_0/A)$, i.e. $\varepsilon_0 + A \cos(\omega t_2) = 0$ and from $t_2\rightarrow t_3$ dynamic phase accumulation
\begin{equation}
    \zeta_2 = \frac{1}{\hbar} \int_{t_2}^{t_3} \sqrt{\varepsilon(t')^2 + \Delta^2} dt',
\end{equation}
then in $t_3$ the anti-crossing is reached again, $t_3 =(2\pi - \arccos(-\varepsilon_0/A))/\omega $. Considering the velocity at the crossings 
\begin{equation}
    v = A\omega \sin(\omega t_2) =  A\omega \sqrt{1 - \left( 
\frac{\varepsilon_0}A \right)^2},
\end{equation}
it can be shown that
\begin{equation}
    P_{LZ}^{\text{double}} = 4 P_{LZ} (1-P_{LZ}) \sin^2(\Phi_S)
    \label{lzs_doble}
\end{equation}
where $P_{LZ} = \exp(-2\pi\frac{\Delta^2}{4\hbar v})$ and $\Phi_S=\phi_S+\zeta_2$ is the Stückelberg phase, which includes the Stokes phase and the dynamic phase.

\begin{figure}[t!]
    \centering
\includegraphics[width=\columnwidth]{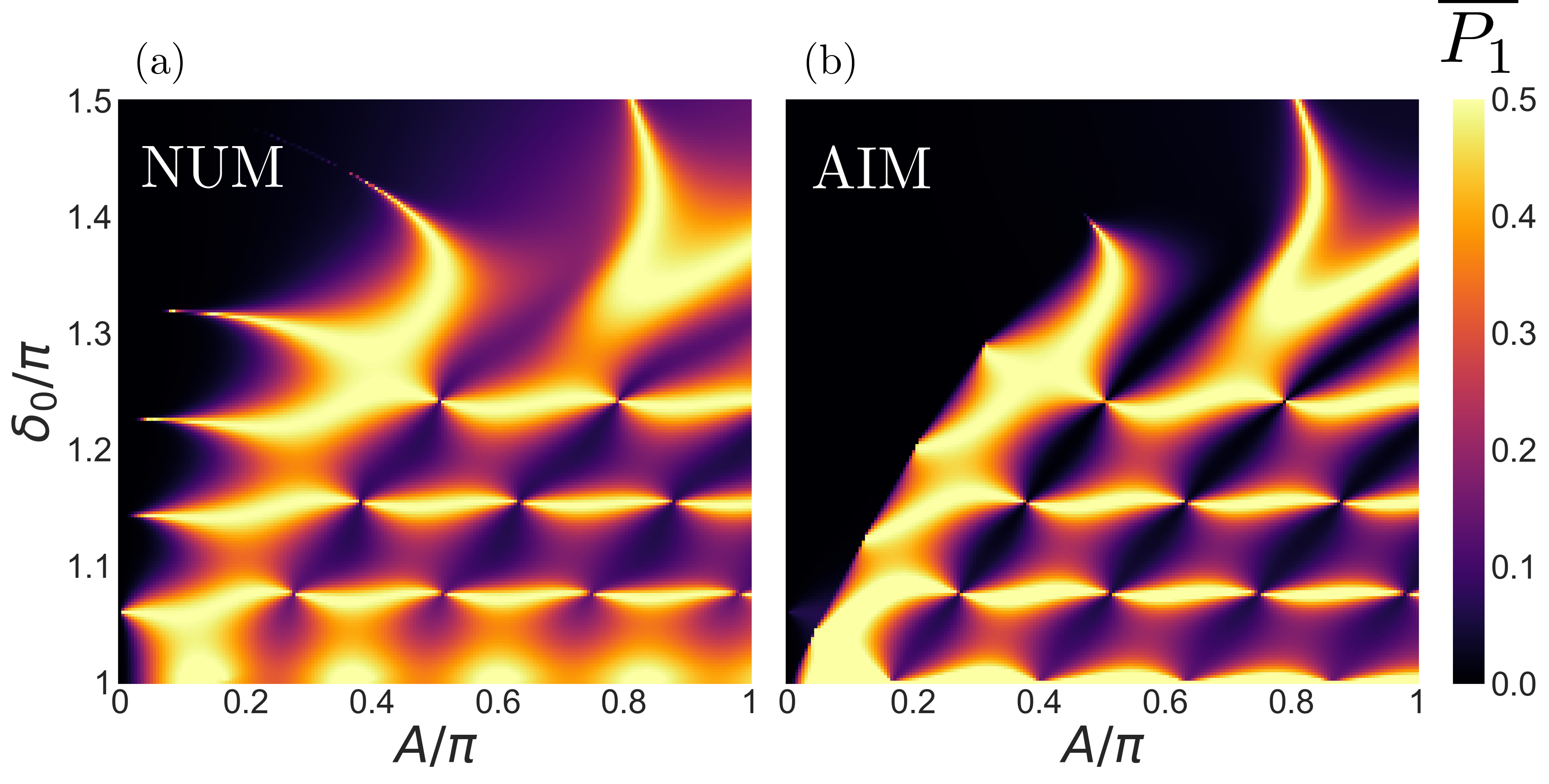}
    \caption{\textbf{LZS spectra for brut-force truncated two-level fluxonium:} Comparison between the numerical result (left) and the analytic result obtained by the AIM approximation (right).}
    \label{fig:LZS_2ls}
\end{figure}

Under periodic driving, the solutions depend on the accumulated phase, effectively behaving like an interferometer. The accumulated phase depends on the parameters of the system, but particularly on $\varepsilon_0$, $A$ and $\omega$. The phases acquired in successive crossings can interfere, this is why the technique of exploring the spectrum using large amplitudes is called Landau-Zener-Stückelberg interferometry. Constructive interference of these phases can lead to so called ``resonances'', where the population completely inverts within each driving period. In Figure \ref{fig:lzs(t)}(b,c) we show two examples of LZS spectra where the time-averaged population of state $|1\rangle$ is simulated as a function of $A$ and $\varepsilon_0$.  The double passage expression van be generalized to $n$ crossings,
\begin{equation}
    P_{LZ}^n =P^{\text{doble}}_{LZ} \frac{\sin^2(n\phi)}{\sin^2(\phi)}
    \label{plz_n_sinpromediar}
\end{equation}
where $\phi$ satisfies
\begin{equation}
    \cos(\phi) = (1- P_{LZ}) \cos(\zeta_+) - P_{LZ} \cos(\zeta_-),
\end{equation}
with
\begin{equation}
    \zeta_+ = \zeta_1 + \zeta_2 + 2\phi_S,\quad \zeta_- = \zeta_1 - \zeta_2.
\end{equation}

Considering the the time-averaged probability and $n \rightarrow\infty$, since $
    \sum_n^\infty \frac{\sin^2(nx)}n = \frac 12$, 
\begin{equation}
    \overline{P}_1=\overline{P}_{LZ} = \frac{P_{LZ}^{\text{doble}}}{2} \frac{1}{\sin^2(\phi)},
\end{equation}
which can be cast as
\begin{equation}
    \overline{P}_{LZ} = \frac 12 \frac{P_{LZ}^{\text{doble}}}{P_{LZ}^{\text{doble}} + |\alpha|^2}.
    \label{lzs_exacto}
\end{equation}
The maximum transition probability $\frac 12$ (resonance condition) occurs when  
\begin{equation}
    |\alpha|^2 = (1 - P_{LZ}) \sin^2(\zeta_+) - P_{LZ} \sin^2(\zeta_-) = 0.
    \label{cond_resonancia}
\end{equation}
In the slow driving regime, $P_{LZ} \approx 0$ and this condition implies
\begin{equation}
    \zeta_1 + \zeta_2 + 2\phi_S = k\pi,
\end{equation}
while in the fast driving regime, $P_{LZ} \approx 1$ and
\begin{equation}
    \zeta_1 - \zeta_2 = k\pi,
    \label{resonancia_rapida}
\end{equation}
here $k \in \mathbb{Z}$. In Fig. \ref{fig:lzs(t)} we show the LZS spectra for (b) fast and (c) slow driving regimes, respectively. In (b) the resonances are equally spaced in the vertical axis $\varepsilon_0/\omega$, which can be derived from the resonance condition for fast driving. From Eq. \eqref{resonancia_rapida}, we obtain that for sufficiently large $A$, such that $\sqrt{(\varepsilon_0 + A\cos(\omega t))^2 + \Delta^2} \approx |\varepsilon_0 + A\cos(\omega t)|$,
\begin{eqnarray*}
    &&\frac{1}{\hbar}\int_{t_2}^{t_3} |\varepsilon_0 + A\cos(\omega t')| dt' \\
    &-&\frac{1}{\hbar}\int_{t_3}^{t_2+\frac{2\pi}\omega} |\varepsilon_0 + A\cos(\omega t')| dt' = k\pi,
\end{eqnarray*}
and since $t_3$ is such that $\varepsilon_0 + A\cos(\omega t) = 0$, there is a minus sign, thus
\begin{equation}
    \frac{1}{\hbar}\int_{t_2}^{t_2 + \frac{2\pi}{\omega}} (\varepsilon_0 + A\cos(\omega t')) dt' = k\pi,
\end{equation}
where the time-average of the $\cos(\omega t')$ vanishes, then 
\begin{equation}
    \varepsilon_0 = k\hbar\omega,
\end{equation}
which is the condition observed in Fig. \ref{fig:lzs(t)}(b).

In the fast driving regime $ A\hbar\omega \gg \Delta^2$, an analytic expression can be found for $\overline{P}_{LZ}(\varepsilon_0,A)$ using a RWA,
\begin{equation}
    \{|0\rangle, |1\rangle\} \longrightarrow \{U(t)|0\rangle, U(t) |1\rangle\} = \{ |0'\rangle, |1'\rangle\},
\end{equation}
such that the Hamiltonian becomes time-independent. For the Hamiltonian in Eq. (\eqref{h_lzs(t)}), we take
\begin{equation}
    U(t) = \text{exp}\left( i f(t) \sigma_z  \right),
\end{equation}
and obtain
\begin{eqnarray*}
    H'(t) &=& \frac 12 (\varepsilon_0 + A\cos(\omega t) - 2  f'(t))\sigma_z \nonumber \\
    &+& \frac{\Delta}2 \left( \frac{e^{2if(t)}  + e^{-2if(t)}}2 \sigma_x+  \frac{e^{2if(t)} - e^{-2if(t)}}{2i} \sigma_y\right),
    \label{rwa_1}
\end{eqnarray*}  
where choosing
\begin{equation}
    f(t) = \frac A{2\hbar} \int_0^t \cos(\omega t') dt' = \frac A{2\hbar\omega} \sin(\omega t),
\end{equation}
and expanding $e^{\pm 2if(t)}$ in terms of Bessel functions of the first kind using Jacobi-Anger expansion we get
\begin{eqnarray*}
    e^{ix\sin(c)} &=& \sum_{n=-\infty}^{\infty} J_n(x) e^{inc}\\
    \Rightarrow e^{\pm i \frac A{\hbar\omega} \sin(\omega t)} &=& \sum_n J_n\left(\frac A{\hbar\omega}\right) e^{in\hbar\omega t},
\end{eqnarray*}
where the terms $n\neq 0$ evolve with fast oscillating exponential terms which average to zero, then $e^{\pm i \frac A{\hbar\omega}\sin(\omega t)} \approx J_0\left(\frac A{\hbar\omega}\right)$, and
\begin{equation}
    H_{\text{RWA}}(t) = \frac {\varepsilon_0}2 \sigma_z + \frac{\Delta}{2} J_0\left( \frac{A}{\hbar\omega} \right)\sigma_x
\end{equation}
with eigenergies $\pm\frac 12 \hbar\Omega_0 = \pm\frac 12 \sqrt{\varepsilon_0^2 + \left( \Delta J_0(\frac A{\hbar\omega}) \right)^2}$. 

Starting at $|0\rangle$ at $t=0$,
\begin{equation}
    P_{01}(t) = \langle 1|e^{-iH_{\text{RWA}}t/\hbar} |0\rangle,
\end{equation}
thus 
\begin{equation}
    P_{01}(t) = \frac 12  \frac{\left(\Delta J_0\left( \frac A{\hbar\omega} \right)\right)^2}{(\hbar\Omega_0)^2} (1- \cos(\Omega_0 t)),
\end{equation}
and therefore, the time-average is
\begin{equation}
    \overline{P}_{01}^0 = \frac 12  \frac{\left(\Delta J_0\left( \frac A{\hbar\omega} \right)\right)^2}{(\hbar\Omega_0)^2},
    \label{Pprom_0}
\end{equation}
where the maximum probability is $\overline{P}_{01}^0 = \frac{1}{2}$. The regions is the LZS spectrum where this value is reached as a function of $(\varepsilon_0, A)$ correspond to the LZS interference resonances. According to Eq. (\eqref{Pprom_0}),  $\overline{P}_{01}^0$ is maximal close to $\varepsilon_0\approx0$, with a modulation on amplitude given by the Bessel function $J_0(A/{\hbar\omega})$. For each resonance, $\varepsilon_0 = k\hbar\omega$, a similar RWA can be performed using $U_k(t)$
\begin{equation}
    U_k(t) = \exp\left( \frac 12 k\omega t\right),
\end{equation}
in which case, 
\begin{equation}
    H_{\text{RWA}}^k = \frac{\varepsilon_0 - k\hbar\omega}2 + \frac{\Delta}2 J_k \left( \frac A{\hbar\omega}\right) \sigma_x,
\end{equation}
where there is a shift $k\hbar\omega$ in $\sigma_z$ and the Bessel function of order $k$. The LZS spectrum is described by
\begin{equation}
    \overline{P}_{01}(\varepsilon_0,A,\omega) = \frac 12  \sum_{k = -\infty}^\infty \frac{\left(\Delta J_k\left( \frac A{\hbar\omega} \right)\right)^2}{(\varepsilon_0 - k\hbar\omega)^2 + \left(\Delta J_k\left( \frac A{\hbar\omega} \right)\right)^2},
    \label{lzs_rapido_bessels}
\end{equation}
which is very good agreement with the spectrum in Fig. \ref{fig:lzs(t)}(b).

\section{LZS for fluxonium truncated to $N_{\text{levels}}=$2,3,4 and 5}
\label{ap:LZS_2ls}

\begin{figure*}[t!]
    \centering
\includegraphics[width=1.9\columnwidth]{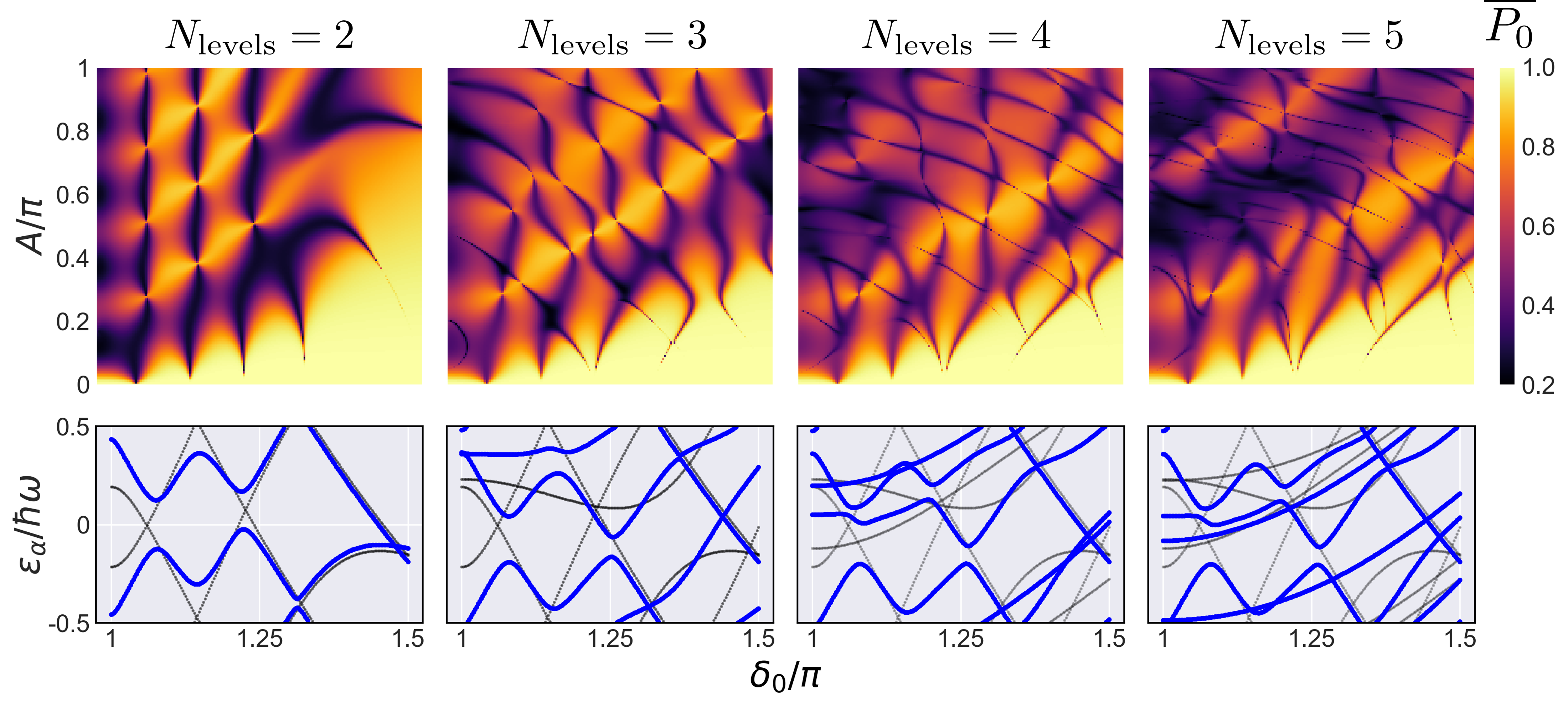}
    \caption{\textbf{LZS spectra for brut-force truncation of the fluxonium at different $N_{\text{level}}$:} (top row) Numerical result of the time-average population of the ground state $\overline{P_0}$ as a function of the drive amplitude $A$ and the phase offset $\delta_0$ obtained from the integration of Schrödinger equation with a truncated Hamiltonian at different $N_{\text{level}}$. The rotation of the axis  of the LZS spectra simplifies the comparison with the corresponding quasienergies presented in the bottom row for $A=0$ (grey) and $A=0.2\pi$ (blue).}
    \label{fig:LZS_3ls}
\end{figure*}

We have discussed in the main text the LZS spectrum obtained from the numerical integration of the Schrödinger equation using a brut-force truncated Hamiltonian with only two-levels (see Fig. \ref{fig:fig3}). Moreover, using an effective LZS-like model and the AIM developed in App. \ref{ap:LZS_classic}, we provide a comparison between the numerical and analytical results in a large range of driving amplitudes in Fig. \ref{fig:LZS_2ls}. The relative magnitudes of the matrix elements $\delta_x$ and $\delta_z$ determine the effect of the driving. When $\delta_z$ dominates ($\delta_z \gg \delta_x$), the drive acts as a LZS drive, effectively proportional to $\sigma_z$. Conversely, when $\delta_x$ dominates, the drive behaves like a Rabi drive, effectively proportional to $\sigma_x$. The AIM for the effective LZS-like Hamiltonian provides a fair agreement for large enough $A$, but it fails to describe the intermediate drive velocity regime and the parameter region where the amplitude is not large enough to reach the anti-crossing, therefore it misses the sharp hook-like features. 

As a first  attempt to understand the complicated spectrum shown in Fig. \ref{fig:fig2} we tried a sequential approach truncating at different $N_{\text{levels}}$ and then looking for superimposed effective two-level systems ($|0\rangle \leftrightarrow |1\rangle$, $|0\rangle \leftrightarrow |2\rangle$, $|1\rangle \leftrightarrow |2\rangle$, etc). 
However, the effect of considering more and more levels cannot be accounted for in this way. In the top panels of 
Fig. \ref{fig:LZS_3ls}, the time-average population of the ground state $\overline{P_0}$, obtained from the integration of the Schrödinger equation with a truncated Hamiltonian at different $N_{\text{level}}= 2,3,4,5$, is plotted, respectively, as a function of the drive amplitude $A$ (vertical axis) and the phase offset $\delta_0$ (horizontal axis). A comparison of the cases $N_{\text{level}}= 2$ and 3 shows the two main effects of the multilevel structure on the LZS spectrum: first, we observe that for $N_{\text{level}}= 3$ the resonance patterns are shifted and twisted; second, new resonance patterns appear (which in some cases seem to cross the previous ones), since they involve transitions to $|2\rangle$. When further considering more levels ($N_{\text{levels}}=4, 5$) the ``two-level'' resonance patterns continues to change and additional resonance patterns appear, with increasing complexity for large $A$ values.

The lower panels of Fig. \ref{fig:LZS_3ls} shows the quasienergies restricted to the first Brillouin zone for $N_{\text{level}}= 2,3,4,5$, respectively, as a function of  $\delta_0$, for  $A=0$ (grey lines) and $A=0.2 \pi$ (blue lines). It can be observed that for low $A$ values, the resonances are mainly located at the crossing of the quasienergies. As a consequence, as more levels are considered, the structure of the spectrum becomes more involved. The effect of the driving amplitude is to open gaps in the
quasienergies spectrum and to couple the different states, shifting the position of the resonances. We can also observe in Fig. \ref{fig:LZS_3ls}, that from $N_{\text{levels}}=$4 to 5 there is only a minor difference at moderate amplitudes, $A\le 0.4\pi$.

\section{LZS for fluxonium: Large amplitude spectrum}
\label{ap:LZS_full_spectra}

\begin{figure}[t!]
\centering
\includegraphics[width=0.5\textwidth]{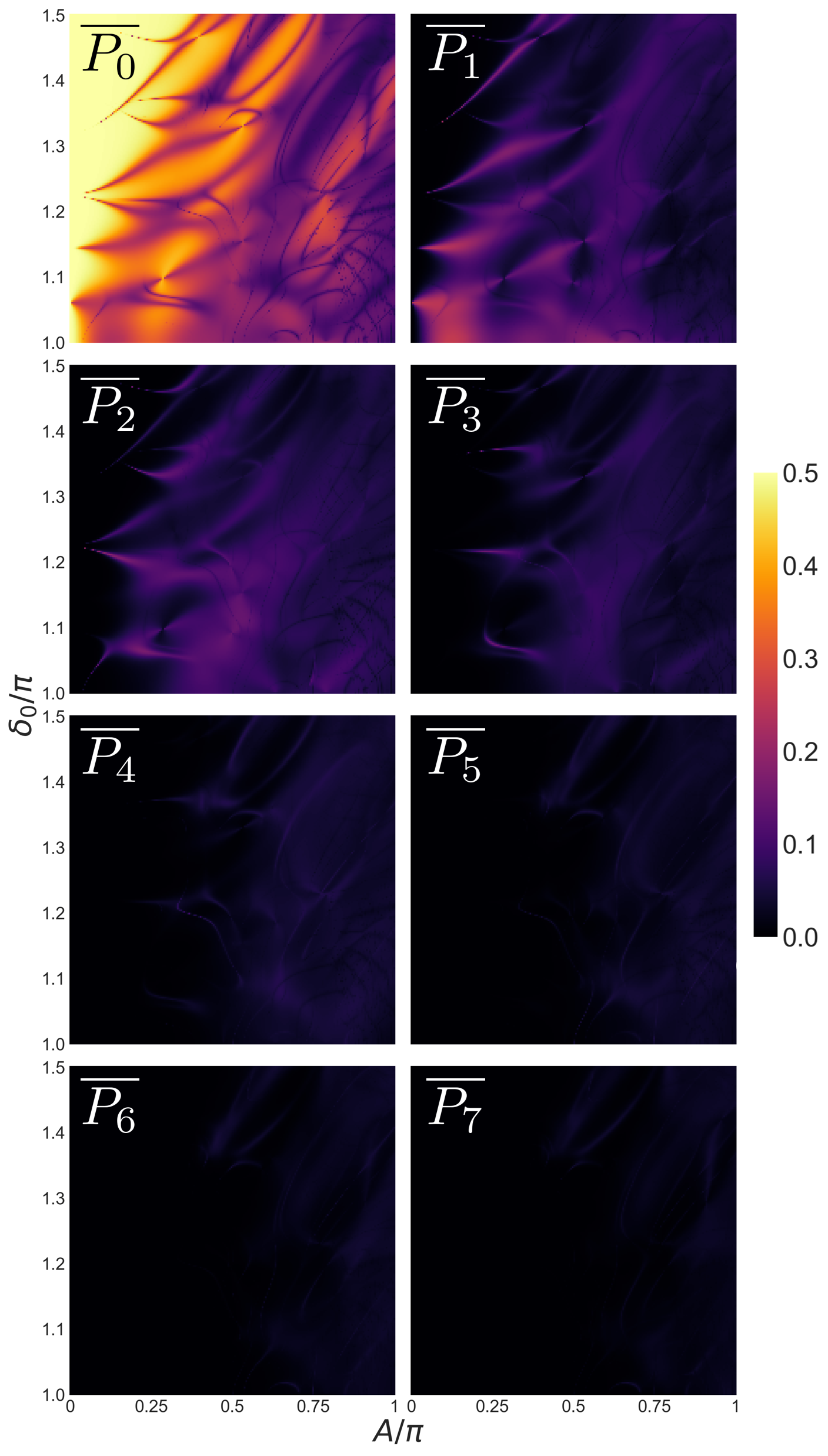}
    \caption{\textbf{LZS Large Amplitude spectrum:} Time-averaged population of the driven fluxonium as a function of drive amplitude $A$ and the external flux $\delta_0$. Averaged population $\overline{P}_i(A,\delta_0)$ with $i=0,7$ in each panel from top to bottom, left to right. The dynamics were simulated using a Hamiltonian truncated to $N_{\text{levels}} = 10$.}
    \label{fig:lzs_muchos}
\end{figure}

Fig. \ref{fig:lzs_muchos} displays the time-averaged populations of the lowest eight energy levels in the periodically driven fluxonium with $A\in [0,\pi]$ and $\delta_0\in[\pi,1.5\pi]$. We have verified that for the parameters used in this work, no significant changes are observed for $N_{\text{levels}}\ge 7$. We have kept $N_{\text{levels}}= 10$ in our simulations of the full multi-level fluxonium. 

Compared to Fig. \ref{fig:fig2} in the Sec. \ref{sec:LZS_num}, here we show the result in much larger amplitude range. The region of large amplitudes is characterized by a significant population of many higher energy levels. The modeling of the expected interferometric spectrum in this regime becomes very hard. As we discuss in the main sections, at lower $A$ exhibits resonance features reminiscent of the two-level LZS spectra and at sufficiently low $A$, the system dynamics effectively reduces to the lowest two energy levels. This regime is therefore most promising for developing low-leakage quantum gates, as shown in Fig. \ref{fig:fig5}.

\section{Floquet Schrieffer-Wolff Approximation}
\label{ap:FSW_app}
We aim to construct a perturbative two-level Hamiltonian that accounts for the presence of upper energy levels and yields a LZS spectrum similar to Fig. \ref{fig:fig2} for low driving amplitudes. For problems where the Hilbert space can be partitioned into two energetically well-separated subspaces, the Schrieffer-Wolff (SW) perturbation theory is a common and effective approach.

Consider a time-independent Hamiltonian of the form
\begin{equation}
    \hat{H} = \hat{H}_0 + \hat{V},
\end{equation}
where $\hat{H}_0$ is a diagonal operator with eigenstates $|i\rangle$ and corresponding energies $\hat{H}_0 |i\rangle = E_i |i\rangle$. The perturbation $\hat{V}$ couples states belonging to two energetically distinct subspaces of $\hat{H}_0$. We can perform a unitary transformation on this Hamiltonian using the operator $\hat{S}$, such that $\hat{H}' = e^{\hat{S}} \hat{H} e^{-\hat{S}}$. Employing the Baker-Campbell-Hausdorff expansion, we obtain
\begin{equation}
    \hat{H}' = \hat{H}_0 + \hat{V} + [\hat{S},\hat{H}_0] + [\hat{S},\hat{V}] + \frac{1}{2} [\hat{S},[\hat{S},\hat{H}_0]] + \mathcal{O}(\hat{S}^3),
\end{equation}
therefore to decouple the two subspaces to first order in the perturbation, we choose $\hat{S}$ such that the off-diagonal elements connecting the subspaces are eliminated. A common choice is to set the terms mixing the subspaces to zero, specifically $\hat{V} + [\hat{S},\hat{H}_0] = 0$. Truncating the expansion to second order in $\hat{S}$ (which is implicitly first order in $\hat{V}$), we are left with a perturbed Hamiltonian
\begin{equation}
    \hat{H}' = \hat{H}_0 + \frac{1}{2} [\hat{S},\hat{V}].
\end{equation}

Projecting this transformed Hamiltonian onto the lower energy subspace results in an effective low-energy Hamiltonian. Given that $\hat{H}_0$ is diagonal, the matrix elements of the transformed Hamiltonian can be explicitly expressed as
\begin{equation}
    \hat{H}'_{ij} = E_i \delta_{i,j} + \frac{1}{2} \sum_{k} V_{ik} V_{kj} \left( \frac{1}{E_i - E_k} + \frac{1}{E_j - E_k} \right)
\end{equation}
where $i,j$ denote states in the low-energy subspace and the summation over $k$ runs over states in the energetically separated (excited) subspace. Here, $\delta_{i,j}$ is the Kronecker delta. 

This result, however, is valid only for a time-independent Hamiltonian. Since our problem is periodic, we can make use of Floquet theory to eliminate this dependence by transforming to an extended Hilbert space known as the Sambe space. In this formalism, the time dependence is effectively encoded within an infinite-dimensional ``photon'' basis, related by a Fourier transform. 
For a Hamiltonian of the form
\begin{eqnarray}
    \hat{H}(t) &=& \hat{H}_0 + 2\hat{V}\cos(\omega t),
\end{eqnarray}
the corresponding Floquet Hamiltonian in the extended Sambe space is given by
\begin{equation}
    H_{F} = 
\left( \begin{array}{ccccc}
\ddots & \hat{V} & 0 & 0 & 0\\ 
V & \hat{H}_0 - \hbar\omega \mathds 1 & \hat{V} & 0 & 0 \\ 
0 & \hat{V} & {H}_0 & \hat{V} & 0 \\ 
0 & 0 & \hat{V} & \hat{H}_0 + \hbar\omega \mathds 1& \hat{V} \\
0 & 0 & 0 & \hat{V} & \ddots \\
\end{array} \right)
\label{sambe}
\end{equation}

The Floquet Hamiltonian matrix exhibits a block tridiagonal structure. The diagonal blocks are equal to the time-independent part of the original Hamiltonian, $\hat{H}_0$, shifted by integer multiples of the driving frequency $\hbar\omega$. These diagonal blocks are coupled only to their nearest neighbors through off-diagonal blocks equal to the interaction term $\hat{V}$. For the fluxonium system, fixing $A$ and $\delta_0$, and truncating to the first three energy levels for illustrative purposes, the Floquet Hamiltonian takes the form
\begin{equation}
    \hat{H}_F = 
\left(
\begin{array}{c|ccc|ccc|c}
\ddots  & \tilde{\delta}_{02} & \tilde{\delta}_{12} & \tilde{\delta}_{22} & 0 & 0 & 0 & \iddots \\ \hline
\tilde{\delta}_{02} & E_0 & 0 & 0 & \tilde{\delta}_{00} & \tilde{\delta}_{01} & \tilde{\delta}_{02} & 0 \\
\tilde{\delta}_{12} & 0 & E_1 & 0 & \tilde{\delta}_{01} & \tilde{\delta}_{11} & \tilde{\delta}_{12} & 0 \\
\tilde{\delta}_{22} & 0 & 0 & E_2 & \tilde{\delta}_{02} & \tilde{\delta}_{12} & \tilde{\delta}_{22} & 0 \\ \hline
0 & \tilde{\delta}_{00} & \tilde{\delta}_{01} & \tilde{\delta}_{02} & E_0+\omega & 0 & 0 & \tilde{\delta}_{00} \\
0 & \tilde{\delta}_{01} & \tilde{\delta}_{11} & \tilde{\delta}_{12} & 0 & E_1+\omega & 0 & \tilde{\delta}_{01} \\
0 & \tilde{\delta}_{02} & \tilde{\delta}_{12} & \tilde{\delta}_{22} & 0 & 0 & E_2+\omega & \tilde{\delta}_{12} \\ \hline
\iddots & 0 & 0 & 0 & \tilde{\delta}_{00} & \tilde{\delta}_{01} & \tilde{\delta}_{02} & \ddots
\end{array}
\right),
\end{equation}
where $E_i\equiv E_i(\delta_0)$ are the eigenenergies of the time-independent fluxonium Hamiltonian, and $\tilde{\delta}_{ij} = (A E_L/2) \langle i (\delta_0)|\hat{\delta}| j (\delta_0)\rangle$. The lines delineate the different photon number blocks. We can decompose $\hat{H}_F$ into elements that act within the qubit subspace $\{|0N\rangle,|1N\rangle\}$, elements that act within the higher energy, non-computational subspace $\{|jN\rangle,|jN\rangle\}$, with $j\ge2$, and elements that couple these two subspaces.
\begin{equation}
\hat{H}_F = \hat{H}_q + \hat{H}_{\overline{q}} + \hat{V}.
\end{equation}
We can apply the SW transformation to this Floquet Hamiltonian to derive an effective 2-qubit Hamiltonian. The matrix elements of this effective Hamiltonian within the $N$-th photon block are given by:
\begin{eqnarray}
    &(\tilde{H}_q)_{ij}^{(N)} = \varepsilon_{iN} \delta_{i,j} + \frac{1}{2} \sum_{k\ge2} V_{ik} V_{kj} \times 
\nonumber \\
&\left[
    \frac{1}{\varepsilon_{iN} - \varepsilon_{kN+1}} + \frac{1}{\varepsilon_{jN} - \varepsilon_{kN+1}}
      \right.
\nonumber \\
&\left. +\frac{1}{\varepsilon_{iN} - \varepsilon_{kN-1}} + \frac{1}{\varepsilon_{jN} - \varepsilon_{kN-1}}
    \right]
    \label{sw}
\end{eqnarray}
where $i,j=0,1$, $V_{ik} = \langle i|\hat{V}|k\rangle$. These dressed elements arise from virtual transitions involving photon absorption or emission $|0 N \rangle \rightarrow |kN\pm 1 \rangle \rightarrow |0N\rangle$ mediated by $\hat{V}$. The energies $\varepsilon_{iN}$ ($\varepsilon_{kN}$) are the eigenvalues of $\hat{H}_q$ ($\hat{H}_{\overline{q}}$), in absence of perturbation. Given that the off-diagonal blocks are proportional to the driving amplitude $A$, for the unperturbed Hamiltonian  $\varepsilon_{iN} = E_i \pm N\hbar\omega$.

There are also second-order processes in $\hat{V}$, which involve block shifts like $|0N\rangle \rightarrow |k N\pm 1\rangle \rightarrow |0N \pm 2\rangle$. The SW transformation introduces terms $\hat{J}$ proportional to $A^2$ that directly couple blocks $N$ and $N \pm 2$. The resulting effective Floquet-Schrieffer-Wolff Hamiltonian is
\begin{equation}
    H_{FSW} =
    \begin{pmatrix}
        \ddots & \hat{V}_q & \hat{J} & 0 & 0 \\
        \hat{V}_q^\dag & \tilde{H}_q - \hbar\omega \mathds1 & \hat{V}_q & \hat{J} & 0 \\
        \hat{J}^\dag & \hat{V}_q^\dag & \tilde{H}_q & \hat{V}_q & \hat{J} \\
        0 & \hat{J}^\dag & \hat{V}_q^\dag & \tilde{H}_q + \hbar\omega \mathds1 & \hat{V}_q \\
        0 & 0 & \hat{J}^\dag & \hat{V}_q^\dag & \ddots \\
    \end{pmatrix}
\end{equation}
where $\tilde{H}_q$ represents the dressed two-level Hamiltonian within a given photon block, with matrix elements given by Eq. \eqref{sw}, $\hat{V}_q = (A E_L/2) \hat{\delta}$ (truncated to two levels), and 
\begin{equation}
    \hat{J} = 
    \begin{pmatrix}
        J_0 & J_{10} \\
        J_{01} & J_1 \\
    \end{pmatrix}
\end{equation}
with
\begin{eqnarray}
\nonumber
J_i &=&\frac 12 \sum_{k\ge2} \tilde{\delta}_{ik}^2 \left[ \frac 1{E_i - E_k - \hbar\omega} + \frac 1{E_i - E_k + \hbar\omega} \right] \\
\nonumber
J_{10} &=& \frac 12\sum_{k\ge2} \tilde{\delta}_{0k}\tilde{\delta}_{k1} \left[ \frac 1{E_0 - E_k + \hbar\omega} + \frac 1{E_1 - E_k - \hbar\omega} \right]\\
\nonumber
J_{01} &=& \frac 12 \sum_{k\ge2} \tilde{\delta}_{0k}\tilde{\delta}_{k1} \left[ \frac 1{E_1 - E_k + \hbar\omega} + \frac 1{E_0 - E_k - \hbar\omega} \right].
\end{eqnarray}

We can return to the time-dependent scheme by realizing that
\begin{equation}
\sum_N |N\rangle \langle N + m| = e^{-im\omega t},
\end{equation}
where $m$ is an arbitrary integer. We finally obtain an effective two-level fluxonium Hamiltonian:
\begin{equation}
    H_{\text{fsw}}(t) = \tilde{H}_q + A E_L \hat{\delta} \cos(\omega t) + \hat{J} e^{-2i\omega t} + \hat{J}^\dag e^{2i\omega t}.
\end{equation}
The term $\hat{J} \propto A^2$ becomes relevant only for values of $A$ for which the low-amplitude approximation breaks down, and thus can be discarded in this regime. We have verified this numerically.

\bibliography{bib/main.bib}

\begin{thebibliography}{56}%
\makeatletter
\providecommand \@ifxundefined [1]{%
 \@ifx{#1\undefined}
}%
\providecommand \@ifnum [1]{%
 \ifnum #1\expandafter \@firstoftwo
 \else \expandafter \@secondoftwo
 \fi
}%
\providecommand \@ifx [1]{%
 \ifx #1\expandafter \@firstoftwo
 \else \expandafter \@secondoftwo
 \fi
}%
\providecommand \natexlab [1]{#1}%
\providecommand \enquote  [1]{``#1''}%
\providecommand \bibnamefont  [1]{#1}%
\providecommand \bibfnamefont [1]{#1}%
\providecommand \citenamefont [1]{#1}%
\providecommand \href@noop [0]{\@secondoftwo}%
\providecommand \href [0]{\begingroup \@sanitize@url \@href}%
\providecommand \@href[1]{\@@startlink{#1}\@@href}%
\providecommand \@@href[1]{\endgroup#1\@@endlink}%
\providecommand \@sanitize@url [0]{\catcode `\\12\catcode `\$12\catcode `\&12\catcode `\#12\catcode `\^12\catcode `\_12\catcode `\%12\relax}%
\providecommand \@@startlink[1]{}%
\providecommand \@@endlink[0]{}%
\providecommand \url  [0]{\begingroup\@sanitize@url \@url }%
\providecommand \@url [1]{\endgroup\@href {#1}{\urlprefix }}%
\providecommand \urlprefix  [0]{URL }%
\providecommand \Eprint [0]{\href }%
\providecommand \doibase [0]{https://doi.org/}%
\providecommand \selectlanguage [0]{\@gobble}%
\providecommand \bibinfo  [0]{\@secondoftwo}%
\providecommand \bibfield  [0]{\@secondoftwo}%
\providecommand \translation [1]{[#1]}%
\providecommand \BibitemOpen [0]{}%
\providecommand \bibitemStop [0]{}%
\providecommand \bibitemNoStop [0]{.\EOS\space}%
\providecommand \EOS [0]{\spacefactor3000\relax}%
\providecommand \BibitemShut  [1]{\csname bibitem#1\endcsname}%
\let\auto@bib@innerbib\@empty
\bibitem [{\citenamefont {Krantz}\ \emph {et~al.}(2019)\citenamefont {Krantz}, \citenamefont {Kjaergaard}, \citenamefont {Yan}, \citenamefont {Orlando}, \citenamefont {Gustavsson},\ and\ \citenamefont {Oliver}}]{Krantz2019}%
  \BibitemOpen
  \bibfield  {author} {\bibinfo {author} {\bibfnamefont {P.}~\bibnamefont {Krantz}}, \bibinfo {author} {\bibfnamefont {M.}~\bibnamefont {Kjaergaard}}, \bibinfo {author} {\bibfnamefont {F.}~\bibnamefont {Yan}}, \bibinfo {author} {\bibfnamefont {T.~P.}\ \bibnamefont {Orlando}}, \bibinfo {author} {\bibfnamefont {S.}~\bibnamefont {Gustavsson}},\ and\ \bibinfo {author} {\bibfnamefont {W.~D.}\ \bibnamefont {Oliver}},\ }\bibfield  {title} {\bibinfo {title} {A quantum engineer’s guide to superconducting qubits},\ }\bibfield  {journal} {\bibinfo  {journal} {Applied Physics Reviews}\ }\textbf {\bibinfo {volume} {6}},\ \href {https://doi.org/10.1063/1.5089550} {10.1063/1.5089550} (\bibinfo {year} {2019})\BibitemShut {NoStop}%
\bibitem [{\citenamefont {Kjaergaard}\ \emph {et~al.}(2020)\citenamefont {Kjaergaard}, \citenamefont {Schwartz}, \citenamefont {Braum{"u}ller}, \citenamefont {Krantz}, \citenamefont {Wang}, \citenamefont {Gustavsson},\ and\ \citenamefont {Oliver}}]{kjaergaard_2020}%
  \BibitemOpen
  \bibfield  {author} {\bibinfo {author} {\bibfnamefont {M.}~\bibnamefont {Kjaergaard}}, \bibinfo {author} {\bibfnamefont {M.}~\bibnamefont {Schwartz}}, \bibinfo {author} {\bibfnamefont {J.}~\bibnamefont {Braum{"u}ller}}, \bibinfo {author} {\bibfnamefont {P.}~\bibnamefont {Krantz}}, \bibinfo {author} {\bibfnamefont {J.~I.-J.}\ \bibnamefont {Wang}}, \bibinfo {author} {\bibfnamefont {S.}~\bibnamefont {Gustavsson}},\ and\ \bibinfo {author} {\bibfnamefont {W.}~\bibnamefont {Oliver}},\ }\bibfield  {title} {\bibinfo {title} {Superconducting qubits: Current state of play},\ }\href@noop {} {\bibfield  {journal} {\bibinfo  {journal} {Annual Review of Condensed Matter Physics}\ }\textbf {\bibinfo {volume} {11}},\ \bibinfo {pages} {369} (\bibinfo {year} {2020})}\BibitemShut {NoStop}%
\bibitem [{\citenamefont {Koch}\ \emph {et~al.}(2007)\citenamefont {Koch}, \citenamefont {Yu}, \citenamefont {Gambetta}, \citenamefont {Houck}, \citenamefont {Schuster}, \citenamefont {Majer}, \citenamefont {Blais}, \citenamefont {Devoret}, \citenamefont {Girvin},\ and\ \citenamefont {Schoelkopf}}]{Koch2007}%
  \BibitemOpen
  \bibfield  {author} {\bibinfo {author} {\bibfnamefont {J.}~\bibnamefont {Koch}}, \bibinfo {author} {\bibfnamefont {T.~M.}\ \bibnamefont {Yu}}, \bibinfo {author} {\bibfnamefont {J.}~\bibnamefont {Gambetta}}, \bibinfo {author} {\bibfnamefont {A.~A.}\ \bibnamefont {Houck}}, \bibinfo {author} {\bibfnamefont {D.~I.}\ \bibnamefont {Schuster}}, \bibinfo {author} {\bibfnamefont {J.}~\bibnamefont {Majer}}, \bibinfo {author} {\bibfnamefont {A.}~\bibnamefont {Blais}}, \bibinfo {author} {\bibfnamefont {M.~H.}\ \bibnamefont {Devoret}}, \bibinfo {author} {\bibfnamefont {S.~M.}\ \bibnamefont {Girvin}},\ and\ \bibinfo {author} {\bibfnamefont {R.~J.}\ \bibnamefont {Schoelkopf}},\ }\bibfield  {title} {\bibinfo {title} {{Charge-insensitive qubit design derived from the Cooper pair box}},\ }\href {https://doi.org/10.1103/PhysRevA.76.042319} {\bibfield  {journal} {\bibinfo  {journal} {Phys. Rev. A}\ }\textbf {\bibinfo {volume} {76}},\ \bibinfo {pages} {042319} (\bibinfo {year} {2007})}\BibitemShut {NoStop}%
\bibitem [{\citenamefont {Manucharyan}\ \emph {et~al.}(2009)\citenamefont {Manucharyan}, \citenamefont {Koch}, \citenamefont {Glazman},\ and\ \citenamefont {Devoret}}]{Manucharyan2009}%
  \BibitemOpen
  \bibfield  {author} {\bibinfo {author} {\bibfnamefont {V.~E.}\ \bibnamefont {Manucharyan}}, \bibinfo {author} {\bibfnamefont {J.}~\bibnamefont {Koch}}, \bibinfo {author} {\bibfnamefont {L.~I.}\ \bibnamefont {Glazman}},\ and\ \bibinfo {author} {\bibfnamefont {M.~H.}\ \bibnamefont {Devoret}},\ }\bibfield  {title} {\bibinfo {title} {{Fluxonium: Single Cooper-Pair Circuit Free of Charge Offsets}},\ }\href {https://doi.org/10.1126/science.1175552} {\bibfield  {journal} {\bibinfo  {journal} {Science}\ }\textbf {\bibinfo {volume} {326}},\ \bibinfo {pages} {113–116} (\bibinfo {year} {2009})}\BibitemShut {NoStop}%
\bibitem [{\citenamefont {Koch}\ \emph {et~al.}(2009)\citenamefont {Koch}, \citenamefont {Manucharyan}, \citenamefont {Devoret},\ and\ \citenamefont {Glazman}}]{Koch2009}%
  \BibitemOpen
  \bibfield  {author} {\bibinfo {author} {\bibfnamefont {J.}~\bibnamefont {Koch}}, \bibinfo {author} {\bibfnamefont {V.}~\bibnamefont {Manucharyan}}, \bibinfo {author} {\bibfnamefont {M.~H.}\ \bibnamefont {Devoret}},\ and\ \bibinfo {author} {\bibfnamefont {L.~I.}\ \bibnamefont {Glazman}},\ }\bibfield  {title} {\bibinfo {title} {{Charging Effects in the Inductively Shunted Josephson Junction}},\ }\href {https://doi.org/10.1103/PhysRevLett.103.217004} {\bibfield  {journal} {\bibinfo  {journal} {Phys. Rev. Lett.}\ }\textbf {\bibinfo {volume} {103}},\ \bibinfo {pages} {217004} (\bibinfo {year} {2009})}\BibitemShut {NoStop}%
\bibitem [{\citenamefont {Pop}\ \emph {et~al.}(2014)\citenamefont {Pop}, \citenamefont {Geerlings}, \citenamefont {Catelani}, \citenamefont {Schoelkopf}, \citenamefont {Glazman},\ and\ \citenamefont {Devoret}}]{Pop2014}%
  \BibitemOpen
  \bibfield  {author} {\bibinfo {author} {\bibfnamefont {I.~M.}\ \bibnamefont {Pop}}, \bibinfo {author} {\bibfnamefont {K.}~\bibnamefont {Geerlings}}, \bibinfo {author} {\bibfnamefont {G.}~\bibnamefont {Catelani}}, \bibinfo {author} {\bibfnamefont {R.~J.}\ \bibnamefont {Schoelkopf}}, \bibinfo {author} {\bibfnamefont {L.~I.}\ \bibnamefont {Glazman}},\ and\ \bibinfo {author} {\bibfnamefont {M.~H.}\ \bibnamefont {Devoret}},\ }\bibfield  {title} {\bibinfo {title} {Coherent suppression of electromagnetic dissipation due to superconducting quasiparticles},\ }\href {https://doi.org/10.1038/nature13017} {\bibfield  {journal} {\bibinfo  {journal} {Nature}\ }\textbf {\bibinfo {volume} {508}},\ \bibinfo {pages} {369–372} (\bibinfo {year} {2014})}\BibitemShut {NoStop}%
\bibitem [{\citenamefont {Nguyen}\ \emph {et~al.}(2019)\citenamefont {Nguyen}, \citenamefont {Lin}, \citenamefont {Somoroff}, \citenamefont {Mencia}, \citenamefont {Grabon},\ and\ \citenamefont {Manucharyan}}]{nguyen_2019}%
  \BibitemOpen
  \bibfield  {author} {\bibinfo {author} {\bibfnamefont {L.~B.}\ \bibnamefont {Nguyen}}, \bibinfo {author} {\bibfnamefont {Y.-H.}\ \bibnamefont {Lin}}, \bibinfo {author} {\bibfnamefont {A.}~\bibnamefont {Somoroff}}, \bibinfo {author} {\bibfnamefont {R.}~\bibnamefont {Mencia}}, \bibinfo {author} {\bibfnamefont {N.}~\bibnamefont {Grabon}},\ and\ \bibinfo {author} {\bibfnamefont {V.~E.}\ \bibnamefont {Manucharyan}},\ }\bibfield  {title} {\bibinfo {title} {High-coherence fluxonium qubit},\ }\href@noop {} {\bibfield  {journal} {\bibinfo  {journal} {Physical Review X}\ }\textbf {\bibinfo {volume} {9}},\ \bibinfo {pages} {041041} (\bibinfo {year} {2019})}\BibitemShut {NoStop}%
\bibitem [{\citenamefont {Gyenis}\ \emph {et~al.}(2021)\citenamefont {Gyenis}, \citenamefont {Di~Paolo}, \citenamefont {Koch}, \citenamefont {Blais}, \citenamefont {Houck},\ and\ \citenamefont {Schuster}}]{gyenis2021}%
  \BibitemOpen
  \bibfield  {author} {\bibinfo {author} {\bibfnamefont {A.}~\bibnamefont {Gyenis}}, \bibinfo {author} {\bibfnamefont {A.}~\bibnamefont {Di~Paolo}}, \bibinfo {author} {\bibfnamefont {J.}~\bibnamefont {Koch}}, \bibinfo {author} {\bibfnamefont {A.}~\bibnamefont {Blais}}, \bibinfo {author} {\bibfnamefont {A.~A.}\ \bibnamefont {Houck}},\ and\ \bibinfo {author} {\bibfnamefont {D.~I.}\ \bibnamefont {Schuster}},\ }\bibfield  {title} {\bibinfo {title} {{Moving beyond the Transmon: Noise-Protected Superconducting Quantum Circuits}},\ }\href {https://doi.org/10.1103/PRXQuantum.2.030101} {\bibfield  {journal} {\bibinfo  {journal} {PRX Quantum}\ }\textbf {\bibinfo {volume} {2}},\ \bibinfo {pages} {030101} (\bibinfo {year} {2021})}\BibitemShut {NoStop}%
\bibitem [{\citenamefont {Bao}\ \emph {et~al.}(2022)\citenamefont {Bao}, \citenamefont {Deng}, \citenamefont {Ding}, \citenamefont {Gao}, \citenamefont {Gao}, \citenamefont {Huang}, \citenamefont {Jiang}, \citenamefont {Ku}, \citenamefont {Li}, \citenamefont {Ma}, \citenamefont {Ni}, \citenamefont {Qin}, \citenamefont {Song}, \citenamefont {Sun}, \citenamefont {Tang}, \citenamefont {Wang}, \citenamefont {Wu}, \citenamefont {Xia}, \citenamefont {Yu}, \citenamefont {Zhang}, \citenamefont {Zhang}, \citenamefont {Zhang}, \citenamefont {Zhou}, \citenamefont {Zhu}, \citenamefont {Shi}, \citenamefont {Chen}, \citenamefont {Zhao},\ and\ \citenamefont {Deng}}]{bao_2022}%
  \BibitemOpen
  \bibfield  {author} {\bibinfo {author} {\bibfnamefont {F.}~\bibnamefont {Bao}}, \bibinfo {author} {\bibfnamefont {H.}~\bibnamefont {Deng}}, \bibinfo {author} {\bibfnamefont {D.}~\bibnamefont {Ding}}, \bibinfo {author} {\bibfnamefont {R.}~\bibnamefont {Gao}}, \bibinfo {author} {\bibfnamefont {X.}~\bibnamefont {Gao}}, \bibinfo {author} {\bibfnamefont {C.}~\bibnamefont {Huang}}, \bibinfo {author} {\bibfnamefont {X.}~\bibnamefont {Jiang}}, \bibinfo {author} {\bibfnamefont {H.-S.}\ \bibnamefont {Ku}}, \bibinfo {author} {\bibfnamefont {Z.}~\bibnamefont {Li}}, \bibinfo {author} {\bibfnamefont {X.}~\bibnamefont {Ma}}, \bibinfo {author} {\bibfnamefont {X.}~\bibnamefont {Ni}}, \bibinfo {author} {\bibfnamefont {J.}~\bibnamefont {Qin}}, \bibinfo {author} {\bibfnamefont {Z.}~\bibnamefont {Song}}, \bibinfo {author} {\bibfnamefont {H.}~\bibnamefont {Sun}}, \bibinfo {author} {\bibfnamefont {C.}~\bibnamefont {Tang}}, \bibinfo {author} {\bibfnamefont {T.}~\bibnamefont {Wang}}, \bibinfo {author} {\bibfnamefont
  {F.}~\bibnamefont {Wu}}, \bibinfo {author} {\bibfnamefont {T.}~\bibnamefont {Xia}}, \bibinfo {author} {\bibfnamefont {W.}~\bibnamefont {Yu}}, \bibinfo {author} {\bibfnamefont {F.}~\bibnamefont {Zhang}}, \bibinfo {author} {\bibfnamefont {G.}~\bibnamefont {Zhang}}, \bibinfo {author} {\bibfnamefont {X.}~\bibnamefont {Zhang}}, \bibinfo {author} {\bibfnamefont {J.}~\bibnamefont {Zhou}}, \bibinfo {author} {\bibfnamefont {X.}~\bibnamefont {Zhu}}, \bibinfo {author} {\bibfnamefont {Y.}~\bibnamefont {Shi}}, \bibinfo {author} {\bibfnamefont {J.}~\bibnamefont {Chen}}, \bibinfo {author} {\bibfnamefont {H.-H.}\ \bibnamefont {Zhao}},\ and\ \bibinfo {author} {\bibfnamefont {C.}~\bibnamefont {Deng}},\ }\bibfield  {title} {\bibinfo {title} {Fluxonium: An alternative qubit platform for high-fidelity operations},\ }\href {https://doi.org/10.1103/PhysRevLett.129.010502} {\bibfield  {journal} {\bibinfo  {journal} {Phys. Rev. Lett.}\ }\textbf {\bibinfo {volume} {129}},\ \bibinfo {pages} {010502} (\bibinfo {year}
  {2022})}\BibitemShut {NoStop}%
\bibitem [{\citenamefont {Weiss}\ \emph {et~al.}(2022)\citenamefont {Weiss}, \citenamefont {Zhang}, \citenamefont {Ding}, \citenamefont {Ma}, \citenamefont {Schuster},\ and\ \citenamefont {Koch}}]{weiss_2022}%
  \BibitemOpen
  \bibfield  {author} {\bibinfo {author} {\bibfnamefont {D.~K.}\ \bibnamefont {Weiss}}, \bibinfo {author} {\bibfnamefont {H.}~\bibnamefont {Zhang}}, \bibinfo {author} {\bibfnamefont {C.}~\bibnamefont {Ding}}, \bibinfo {author} {\bibfnamefont {Y.}~\bibnamefont {Ma}}, \bibinfo {author} {\bibfnamefont {D.~I.}\ \bibnamefont {Schuster}},\ and\ \bibinfo {author} {\bibfnamefont {J.}~\bibnamefont {Koch}},\ }\bibfield  {title} {\bibinfo {title} {Fast high-fidelity gates for galvanically-coupled fluxonium qubits using strong flux modulation},\ }\href@noop {} {\bibfield  {journal} {\bibinfo  {journal} {PRX Quantum}\ }\textbf {\bibinfo {volume} {3}},\ \bibinfo {pages} {040336} (\bibinfo {year} {2022})}\BibitemShut {NoStop}%
\bibitem [{\citenamefont {Nguyen}\ \emph {et~al.}(2022)\citenamefont {Nguyen}, \citenamefont {Koolstra}, \citenamefont {Kim}, \citenamefont {Morvan}, \citenamefont {Chistolini}, \citenamefont {Singh}, \citenamefont {Nesterov}, \citenamefont {J\"unger}, \citenamefont {Chen}, \citenamefont {Pedramrazi}, \citenamefont {Mitchell}, \citenamefont {Kreikebaum}, \citenamefont {Puri}, \citenamefont {Santiago},\ and\ \citenamefont {Siddiqi}}]{nguyen_2022}%
  \BibitemOpen
  \bibfield  {author} {\bibinfo {author} {\bibfnamefont {L.~B.}\ \bibnamefont {Nguyen}}, \bibinfo {author} {\bibfnamefont {G.}~\bibnamefont {Koolstra}}, \bibinfo {author} {\bibfnamefont {Y.}~\bibnamefont {Kim}}, \bibinfo {author} {\bibfnamefont {A.}~\bibnamefont {Morvan}}, \bibinfo {author} {\bibfnamefont {T.}~\bibnamefont {Chistolini}}, \bibinfo {author} {\bibfnamefont {S.}~\bibnamefont {Singh}}, \bibinfo {author} {\bibfnamefont {K.~N.}\ \bibnamefont {Nesterov}}, \bibinfo {author} {\bibfnamefont {C.}~\bibnamefont {J\"unger}}, \bibinfo {author} {\bibfnamefont {L.}~\bibnamefont {Chen}}, \bibinfo {author} {\bibfnamefont {Z.}~\bibnamefont {Pedramrazi}}, \bibinfo {author} {\bibfnamefont {B.~K.}\ \bibnamefont {Mitchell}}, \bibinfo {author} {\bibfnamefont {J.~M.}\ \bibnamefont {Kreikebaum}}, \bibinfo {author} {\bibfnamefont {S.}~\bibnamefont {Puri}}, \bibinfo {author} {\bibfnamefont {D.~I.}\ \bibnamefont {Santiago}},\ and\ \bibinfo {author} {\bibfnamefont {I.}~\bibnamefont {Siddiqi}},\ }\bibfield  {title} {\bibinfo
  {title} {Blueprint for a high-performance fluxonium quantum processor},\ }\href {https://doi.org/10.1103/PRXQuantum.3.037001} {\bibfield  {journal} {\bibinfo  {journal} {PRX Quantum}\ }\textbf {\bibinfo {volume} {3}},\ \bibinfo {pages} {037001} (\bibinfo {year} {2022})}\BibitemShut {NoStop}%
\bibitem [{\citenamefont {Ding}\ \emph {et~al.}(2023)\citenamefont {Ding}, \citenamefont {Hays}, \citenamefont {Sung}, \citenamefont {Kannan}, \citenamefont {An}, \citenamefont {Di~Paolo}, \citenamefont {Karamlou}, \citenamefont {Hazard}, \citenamefont {Azar}, \citenamefont {Kim}, \citenamefont {Niedzielski}, \citenamefont {Melville}, \citenamefont {Schwartz}, \citenamefont {Yoder}, \citenamefont {Orlando}, \citenamefont {Gustavsson}, \citenamefont {Grover}, \citenamefont {Serniak},\ and\ \citenamefont {Oliver}}]{ding_2023}%
  \BibitemOpen
  \bibfield  {author} {\bibinfo {author} {\bibfnamefont {L.}~\bibnamefont {Ding}}, \bibinfo {author} {\bibfnamefont {M.}~\bibnamefont {Hays}}, \bibinfo {author} {\bibfnamefont {Y.}~\bibnamefont {Sung}}, \bibinfo {author} {\bibfnamefont {B.}~\bibnamefont {Kannan}}, \bibinfo {author} {\bibfnamefont {J.}~\bibnamefont {An}}, \bibinfo {author} {\bibfnamefont {A.}~\bibnamefont {Di~Paolo}}, \bibinfo {author} {\bibfnamefont {A.~H.}\ \bibnamefont {Karamlou}}, \bibinfo {author} {\bibfnamefont {T.~M.}\ \bibnamefont {Hazard}}, \bibinfo {author} {\bibfnamefont {K.}~\bibnamefont {Azar}}, \bibinfo {author} {\bibfnamefont {D.~K.}\ \bibnamefont {Kim}}, \bibinfo {author} {\bibfnamefont {B.~M.}\ \bibnamefont {Niedzielski}}, \bibinfo {author} {\bibfnamefont {A.}~\bibnamefont {Melville}}, \bibinfo {author} {\bibfnamefont {M.~E.}\ \bibnamefont {Schwartz}}, \bibinfo {author} {\bibfnamefont {J.~L.}\ \bibnamefont {Yoder}}, \bibinfo {author} {\bibfnamefont {T.~P.}\ \bibnamefont {Orlando}}, \bibinfo {author} {\bibfnamefont
  {S.}~\bibnamefont {Gustavsson}}, \bibinfo {author} {\bibfnamefont {J.~A.}\ \bibnamefont {Grover}}, \bibinfo {author} {\bibfnamefont {K.}~\bibnamefont {Serniak}},\ and\ \bibinfo {author} {\bibfnamefont {W.~D.}\ \bibnamefont {Oliver}},\ }\bibfield  {title} {\bibinfo {title} {High-fidelity, frequency-flexible two-qubit fluxonium gates with a transmon coupler},\ }\href {https://doi.org/10.1103/PhysRevX.13.031035} {\bibfield  {journal} {\bibinfo  {journal} {Phys. Rev. X}\ }\textbf {\bibinfo {volume} {13}},\ \bibinfo {pages} {031035} (\bibinfo {year} {2023})}\BibitemShut {NoStop}%
\bibitem [{\citenamefont {Lin}\ \emph {et~al.}(2018)\citenamefont {Lin}, \citenamefont {Nguyen}, \citenamefont {Grabon}, \citenamefont {San~Miguel}, \citenamefont {Pankratova},\ and\ \citenamefont {Manucharyan}}]{lin_2018}%
  \BibitemOpen
  \bibfield  {author} {\bibinfo {author} {\bibfnamefont {Y.-H.}\ \bibnamefont {Lin}}, \bibinfo {author} {\bibfnamefont {L.~B.}\ \bibnamefont {Nguyen}}, \bibinfo {author} {\bibfnamefont {N.}~\bibnamefont {Grabon}}, \bibinfo {author} {\bibfnamefont {J.}~\bibnamefont {San~Miguel}}, \bibinfo {author} {\bibfnamefont {N.}~\bibnamefont {Pankratova}},\ and\ \bibinfo {author} {\bibfnamefont {V.~E.}\ \bibnamefont {Manucharyan}},\ }\bibfield  {title} {\bibinfo {title} {Demonstration of protection of a superconducting qubit from energy decay},\ }\href {https://doi.org/10.1103/PhysRevLett.120.150503} {\bibfield  {journal} {\bibinfo  {journal} {Phys. Rev. Lett.}\ }\textbf {\bibinfo {volume} {120}},\ \bibinfo {pages} {150503} (\bibinfo {year} {2018})}\BibitemShut {NoStop}%
\bibitem [{\citenamefont {Nesterov}\ \emph {et~al.}(2021)\citenamefont {Nesterov}, \citenamefont {Ficheux}, \citenamefont {Manucharyan},\ and\ \citenamefont {Vavilov}}]{nesterov_2021}%
  \BibitemOpen
  \bibfield  {author} {\bibinfo {author} {\bibfnamefont {K.~N.}\ \bibnamefont {Nesterov}}, \bibinfo {author} {\bibfnamefont {Q.}~\bibnamefont {Ficheux}}, \bibinfo {author} {\bibfnamefont {V.~E.}\ \bibnamefont {Manucharyan}},\ and\ \bibinfo {author} {\bibfnamefont {M.~G.}\ \bibnamefont {Vavilov}},\ }\bibfield  {title} {\bibinfo {title} {Proposal for entangling gates on fluxonium qubits via a two-photon transition},\ }\href {https://doi.org/10.1103/PRXQuantum.2.020345} {\bibfield  {journal} {\bibinfo  {journal} {PRX Quantum}\ }\textbf {\bibinfo {volume} {2}},\ \bibinfo {pages} {020345} (\bibinfo {year} {2021})}\BibitemShut {NoStop}%
\bibitem [{\citenamefont {Moskalenko}\ \emph {et~al.}(2021)\citenamefont {Moskalenko}, \citenamefont {Besedin}, \citenamefont {Simakov},\ and\ \citenamefont {Ustinov}}]{moska_2021}%
  \BibitemOpen
  \bibfield  {author} {\bibinfo {author} {\bibfnamefont {I.~N.}\ \bibnamefont {Moskalenko}}, \bibinfo {author} {\bibfnamefont {I.~S.}\ \bibnamefont {Besedin}}, \bibinfo {author} {\bibfnamefont {I.~A.}\ \bibnamefont {Simakov}},\ and\ \bibinfo {author} {\bibfnamefont {A.~V.}\ \bibnamefont {Ustinov}},\ }\bibfield  {title} {\bibinfo {title} {{Tunable coupling scheme for implementing two-qubit gates on fluxonium qubits}},\ }\href {https://doi.org/10.1063/5.0064800} {\bibfield  {journal} {\bibinfo  {journal} {Applied Physics Letters}\ }\textbf {\bibinfo {volume} {119}},\ \bibinfo {pages} {194001} (\bibinfo {year} {2021})}\BibitemShut {NoStop}%
\bibitem [{\citenamefont {Moskalenko}\ \emph {et~al.}(2022)\citenamefont {Moskalenko}, \citenamefont {Simakov}, \citenamefont {Abramov}, \citenamefont {Grigorev}, \citenamefont {Moskalev}, \citenamefont {Pishchimova}, \citenamefont {Smirnov}, \citenamefont {Zikiy}, \citenamefont {Rodionov},\ and\ \citenamefont {Besedin}}]{moska_2022}%
  \BibitemOpen
  \bibfield  {author} {\bibinfo {author} {\bibfnamefont {I.~N.}\ \bibnamefont {Moskalenko}}, \bibinfo {author} {\bibfnamefont {I.~A.}\ \bibnamefont {Simakov}}, \bibinfo {author} {\bibfnamefont {N.~N.}\ \bibnamefont {Abramov}}, \bibinfo {author} {\bibfnamefont {A.~A.}\ \bibnamefont {Grigorev}}, \bibinfo {author} {\bibfnamefont {D.~O.}\ \bibnamefont {Moskalev}}, \bibinfo {author} {\bibfnamefont {A.~A.}\ \bibnamefont {Pishchimova}}, \bibinfo {author} {\bibfnamefont {N.~S.}\ \bibnamefont {Smirnov}}, \bibinfo {author} {\bibfnamefont {E.~V.}\ \bibnamefont {Zikiy}}, \bibinfo {author} {\bibfnamefont {I.~A.}\ \bibnamefont {Rodionov}},\ and\ \bibinfo {author} {\bibfnamefont {I.~S.}\ \bibnamefont {Besedin}},\ }\bibfield  {title} {\bibinfo {title} {High fidelity two-qubit gates on fluxoniums using a tunable coupler},\ }\href {https://doi.org/10.1038/s41534-022-00644-x} {\bibfield  {journal} {\bibinfo  {journal} {npj Quantum Information}\ }\textbf {\bibinfo {volume} {8}},\ \bibinfo {pages} {130} (\bibinfo {year}
  {2022})}\BibitemShut {NoStop}%
\bibitem [{\citenamefont {Maleeva}\ \emph {et~al.}(2018)\citenamefont {Maleeva}, \citenamefont {Gr{\"u}nhaupt}, \citenamefont {Klein}, \citenamefont {Levy-Bertrand}, \citenamefont {Dupre}, \citenamefont {Calvo}, \citenamefont {Valenti}, \citenamefont {Winkel}, \citenamefont {Friedrich}, \citenamefont {Wernsdorfer} \emph {et~al.}}]{maleeva2018circuit}%
  \BibitemOpen
  \bibfield  {author} {\bibinfo {author} {\bibfnamefont {N.}~\bibnamefont {Maleeva}}, \bibinfo {author} {\bibfnamefont {L.}~\bibnamefont {Gr{\"u}nhaupt}}, \bibinfo {author} {\bibfnamefont {T.}~\bibnamefont {Klein}}, \bibinfo {author} {\bibfnamefont {F.}~\bibnamefont {Levy-Bertrand}}, \bibinfo {author} {\bibfnamefont {O.}~\bibnamefont {Dupre}}, \bibinfo {author} {\bibfnamefont {M.}~\bibnamefont {Calvo}}, \bibinfo {author} {\bibfnamefont {F.}~\bibnamefont {Valenti}}, \bibinfo {author} {\bibfnamefont {P.}~\bibnamefont {Winkel}}, \bibinfo {author} {\bibfnamefont {F.}~\bibnamefont {Friedrich}}, \bibinfo {author} {\bibfnamefont {W.}~\bibnamefont {Wernsdorfer}}, \emph {et~al.},\ }\bibfield  {title} {\bibinfo {title} {Circuit quantum electrodynamics of granular aluminum resonators},\ }\href@noop {} {\bibfield  {journal} {\bibinfo  {journal} {Nature communications}\ }\textbf {\bibinfo {volume} {9}},\ \bibinfo {pages} {3889} (\bibinfo {year} {2018})}\BibitemShut {NoStop}%
\bibitem [{\citenamefont {Gr{\"u}nhaupt}\ \emph {et~al.}(2019)\citenamefont {Gr{\"u}nhaupt}, \citenamefont {Spiecker}, \citenamefont {Gusenkova}, \citenamefont {Maleeva}, \citenamefont {Skacel}, \citenamefont {Takmakov}, \citenamefont {Valenti}, \citenamefont {Winkel}, \citenamefont {Rotzinger}, \citenamefont {Wernsdorfer} \emph {et~al.}}]{grunhaupt2019granular}%
  \BibitemOpen
  \bibfield  {author} {\bibinfo {author} {\bibfnamefont {L.}~\bibnamefont {Gr{\"u}nhaupt}}, \bibinfo {author} {\bibfnamefont {M.}~\bibnamefont {Spiecker}}, \bibinfo {author} {\bibfnamefont {D.}~\bibnamefont {Gusenkova}}, \bibinfo {author} {\bibfnamefont {N.}~\bibnamefont {Maleeva}}, \bibinfo {author} {\bibfnamefont {S.~T.}\ \bibnamefont {Skacel}}, \bibinfo {author} {\bibfnamefont {I.}~\bibnamefont {Takmakov}}, \bibinfo {author} {\bibfnamefont {F.}~\bibnamefont {Valenti}}, \bibinfo {author} {\bibfnamefont {P.}~\bibnamefont {Winkel}}, \bibinfo {author} {\bibfnamefont {H.}~\bibnamefont {Rotzinger}}, \bibinfo {author} {\bibfnamefont {W.}~\bibnamefont {Wernsdorfer}}, \emph {et~al.},\ }\bibfield  {title} {\bibinfo {title} {Granular aluminium as a superconducting material for high-impedance quantum circuits},\ }\href@noop {} {\bibfield  {journal} {\bibinfo  {journal} {Nature materials}\ }\textbf {\bibinfo {volume} {18}},\ \bibinfo {pages} {816} (\bibinfo {year} {2019})}\BibitemShut {NoStop}%
\bibitem [{\citenamefont {Levy-Bertrand}\ \emph {et~al.}(2019)\citenamefont {Levy-Bertrand}, \citenamefont {Klein}, \citenamefont {Grenet}, \citenamefont {Dupr{\'e}}, \citenamefont {Beno{\^\i}t}, \citenamefont {Bideaud}, \citenamefont {Bourrion}, \citenamefont {Calvo}, \citenamefont {Catalano}, \citenamefont {Gomez} \emph {et~al.}}]{levy2019electrodynamics}%
  \BibitemOpen
  \bibfield  {author} {\bibinfo {author} {\bibfnamefont {F.}~\bibnamefont {Levy-Bertrand}}, \bibinfo {author} {\bibfnamefont {T.}~\bibnamefont {Klein}}, \bibinfo {author} {\bibfnamefont {T.}~\bibnamefont {Grenet}}, \bibinfo {author} {\bibfnamefont {O.}~\bibnamefont {Dupr{\'e}}}, \bibinfo {author} {\bibfnamefont {A.}~\bibnamefont {Beno{\^\i}t}}, \bibinfo {author} {\bibfnamefont {A.}~\bibnamefont {Bideaud}}, \bibinfo {author} {\bibfnamefont {O.}~\bibnamefont {Bourrion}}, \bibinfo {author} {\bibfnamefont {M.}~\bibnamefont {Calvo}}, \bibinfo {author} {\bibfnamefont {A.}~\bibnamefont {Catalano}}, \bibinfo {author} {\bibfnamefont {A.}~\bibnamefont {Gomez}}, \emph {et~al.},\ }\bibfield  {title} {\bibinfo {title} {Electrodynamics of granular aluminum from superconductor to insulator: Observation of collective superconducting modes},\ }\href@noop {} {\bibfield  {journal} {\bibinfo  {journal} {Physical Review B}\ }\textbf {\bibinfo {volume} {99}},\ \bibinfo {pages} {094506} (\bibinfo {year} {2019})}\BibitemShut
  {NoStop}%
\bibitem [{\citenamefont {Winkel}\ \emph {et~al.}(2020)\citenamefont {Winkel}, \citenamefont {Borisov}, \citenamefont {Gr{\"u}nhaupt}, \citenamefont {Rieger}, \citenamefont {Spiecker}, \citenamefont {Valenti}, \citenamefont {Ustinov}, \citenamefont {Wernsdorfer},\ and\ \citenamefont {Pop}}]{winkel2020implementation}%
  \BibitemOpen
  \bibfield  {author} {\bibinfo {author} {\bibfnamefont {P.}~\bibnamefont {Winkel}}, \bibinfo {author} {\bibfnamefont {K.}~\bibnamefont {Borisov}}, \bibinfo {author} {\bibfnamefont {L.}~\bibnamefont {Gr{\"u}nhaupt}}, \bibinfo {author} {\bibfnamefont {D.}~\bibnamefont {Rieger}}, \bibinfo {author} {\bibfnamefont {M.}~\bibnamefont {Spiecker}}, \bibinfo {author} {\bibfnamefont {F.}~\bibnamefont {Valenti}}, \bibinfo {author} {\bibfnamefont {A.~V.}\ \bibnamefont {Ustinov}}, \bibinfo {author} {\bibfnamefont {W.}~\bibnamefont {Wernsdorfer}},\ and\ \bibinfo {author} {\bibfnamefont {I.~M.}\ \bibnamefont {Pop}},\ }\bibfield  {title} {\bibinfo {title} {Implementation of a transmon qubit using superconducting granular aluminum},\ }\href@noop {} {\bibfield  {journal} {\bibinfo  {journal} {Physical Review X}\ }\textbf {\bibinfo {volume} {10}},\ \bibinfo {pages} {031032} (\bibinfo {year} {2020})}\BibitemShut {NoStop}%
\bibitem [{\citenamefont {Rieger}\ \emph {et~al.}(2023)\citenamefont {Rieger}, \citenamefont {G{\"u}nzler}, \citenamefont {Spiecker}, \citenamefont {Paluch}, \citenamefont {Winkel}, \citenamefont {Hahn}, \citenamefont {Hohmann}, \citenamefont {Bacher}, \citenamefont {Wernsdorfer},\ and\ \citenamefont {Pop}}]{rieger2023granular}%
  \BibitemOpen
  \bibfield  {author} {\bibinfo {author} {\bibfnamefont {D.}~\bibnamefont {Rieger}}, \bibinfo {author} {\bibfnamefont {S.}~\bibnamefont {G{\"u}nzler}}, \bibinfo {author} {\bibfnamefont {M.}~\bibnamefont {Spiecker}}, \bibinfo {author} {\bibfnamefont {P.}~\bibnamefont {Paluch}}, \bibinfo {author} {\bibfnamefont {P.}~\bibnamefont {Winkel}}, \bibinfo {author} {\bibfnamefont {L.}~\bibnamefont {Hahn}}, \bibinfo {author} {\bibfnamefont {J.}~\bibnamefont {Hohmann}}, \bibinfo {author} {\bibfnamefont {A.}~\bibnamefont {Bacher}}, \bibinfo {author} {\bibfnamefont {W.}~\bibnamefont {Wernsdorfer}},\ and\ \bibinfo {author} {\bibfnamefont {I.}~\bibnamefont {Pop}},\ }\bibfield  {title} {\bibinfo {title} {Granular aluminium nanojunction fluxonium qubit},\ }\href@noop {} {\bibfield  {journal} {\bibinfo  {journal} {Nature Materials}\ }\textbf {\bibinfo {volume} {22}},\ \bibinfo {pages} {194} (\bibinfo {year} {2023})}\BibitemShut {NoStop}%
\bibitem [{\citenamefont {Zhang}\ \emph {et~al.}(2021)\citenamefont {Zhang}, \citenamefont {Chakram}, \citenamefont {Roy}, \citenamefont {Earnest}, \citenamefont {Lu}, \citenamefont {Huang}, \citenamefont {Weiss}, \citenamefont {Koch},\ and\ \citenamefont {Schuster}}]{zhang_2021}%
  \BibitemOpen
  \bibfield  {author} {\bibinfo {author} {\bibfnamefont {H.}~\bibnamefont {Zhang}}, \bibinfo {author} {\bibfnamefont {S.}~\bibnamefont {Chakram}}, \bibinfo {author} {\bibfnamefont {T.}~\bibnamefont {Roy}}, \bibinfo {author} {\bibfnamefont {N.}~\bibnamefont {Earnest}}, \bibinfo {author} {\bibfnamefont {Y.}~\bibnamefont {Lu}}, \bibinfo {author} {\bibfnamefont {Z.}~\bibnamefont {Huang}}, \bibinfo {author} {\bibfnamefont {D.~K.}\ \bibnamefont {Weiss}}, \bibinfo {author} {\bibfnamefont {J.}~\bibnamefont {Koch}},\ and\ \bibinfo {author} {\bibfnamefont {D.~I.}\ \bibnamefont {Schuster}},\ }\bibfield  {title} {\bibinfo {title} {Universal fast-flux control of a coherent, low-frequency qubit},\ }\href {https://doi.org/10.1103/PhysRevX.11.011010} {\bibfield  {journal} {\bibinfo  {journal} {Phys. Rev. X}\ }\textbf {\bibinfo {volume} {11}},\ \bibinfo {pages} {011010} (\bibinfo {year} {2021})}\BibitemShut {NoStop}%
\bibitem [{\citenamefont {Somoroff}\ \emph {et~al.}(2023)\citenamefont {Somoroff}, \citenamefont {Ficheux}, \citenamefont {Mencia}, \citenamefont {Xiong}, \citenamefont {Kuzmin},\ and\ \citenamefont {Manucharyan}}]{somoroff_2023}%
  \BibitemOpen
  \bibfield  {author} {\bibinfo {author} {\bibfnamefont {A.}~\bibnamefont {Somoroff}}, \bibinfo {author} {\bibfnamefont {Q.}~\bibnamefont {Ficheux}}, \bibinfo {author} {\bibfnamefont {R.~A.}\ \bibnamefont {Mencia}}, \bibinfo {author} {\bibfnamefont {H.}~\bibnamefont {Xiong}}, \bibinfo {author} {\bibfnamefont {R.}~\bibnamefont {Kuzmin}},\ and\ \bibinfo {author} {\bibfnamefont {V.~E.}\ \bibnamefont {Manucharyan}},\ }\bibfield  {title} {\bibinfo {title} {Millisecond coherence in a superconducting qubit},\ }\href {https://doi.org/10.1103/PhysRevLett.130.267001} {\bibfield  {journal} {\bibinfo  {journal} {Phys. Rev. Lett.}\ }\textbf {\bibinfo {volume} {130}},\ \bibinfo {pages} {267001} (\bibinfo {year} {2023})}\BibitemShut {NoStop}%
\bibitem [{\citenamefont {Rower}\ \emph {et~al.}(2024)\citenamefont {Rower}, \citenamefont {Ding}, \citenamefont {Zhang}, \citenamefont {Hays}, \citenamefont {An}, \citenamefont {Harrington}, \citenamefont {Rosen}, \citenamefont {Gertler}, \citenamefont {Hazard}, \citenamefont {Niedzielski}, \citenamefont {Schwartz}, \citenamefont {Gustavsson}, \citenamefont {Serniak}, \citenamefont {Grover},\ and\ \citenamefont {Oliver}}]{rower_2024}%
  \BibitemOpen
  \bibfield  {author} {\bibinfo {author} {\bibfnamefont {D.~A.}\ \bibnamefont {Rower}}, \bibinfo {author} {\bibfnamefont {L.}~\bibnamefont {Ding}}, \bibinfo {author} {\bibfnamefont {H.}~\bibnamefont {Zhang}}, \bibinfo {author} {\bibfnamefont {M.}~\bibnamefont {Hays}}, \bibinfo {author} {\bibfnamefont {J.}~\bibnamefont {An}}, \bibinfo {author} {\bibfnamefont {P.~M.}\ \bibnamefont {Harrington}}, \bibinfo {author} {\bibfnamefont {I.~T.}\ \bibnamefont {Rosen}}, \bibinfo {author} {\bibfnamefont {J.~M.}\ \bibnamefont {Gertler}}, \bibinfo {author} {\bibfnamefont {T.~M.}\ \bibnamefont {Hazard}}, \bibinfo {author} {\bibfnamefont {B.~M.}\ \bibnamefont {Niedzielski}}, \bibinfo {author} {\bibfnamefont {M.~E.}\ \bibnamefont {Schwartz}}, \bibinfo {author} {\bibfnamefont {S.}~\bibnamefont {Gustavsson}}, \bibinfo {author} {\bibfnamefont {K.}~\bibnamefont {Serniak}}, \bibinfo {author} {\bibfnamefont {J.~A.}\ \bibnamefont {Grover}},\ and\ \bibinfo {author} {\bibfnamefont {W.~D.}\ \bibnamefont {Oliver}},\ }\bibfield  {title}
  {\bibinfo {title} {Suppressing counter-rotating errors for fast single-qubit gates with fluxonium},\ }\href {https://doi.org/10.1103/PRXQuantum.5.040342} {\bibfield  {journal} {\bibinfo  {journal} {PRX Quantum}\ }\textbf {\bibinfo {volume} {5}},\ \bibinfo {pages} {040342} (\bibinfo {year} {2024})}\BibitemShut {NoStop}%
\bibitem [{\citenamefont {Shevchenko}\ \emph {et~al.}(2010)\citenamefont {Shevchenko}, \citenamefont {Ashhab},\ and\ \citenamefont {Nori}}]{shevchenko_2010}%
  \BibitemOpen
  \bibfield  {author} {\bibinfo {author} {\bibfnamefont {S.}~\bibnamefont {Shevchenko}}, \bibinfo {author} {\bibfnamefont {S.}~\bibnamefont {Ashhab}},\ and\ \bibinfo {author} {\bibfnamefont {F.}~\bibnamefont {Nori}},\ }\bibfield  {title} {\bibinfo {title} {Landau–zener–stückelberg interferometry},\ }\href {https://doi.org/https://doi.org/10.1016/j.physrep.2010.03.002} {\bibfield  {journal} {\bibinfo  {journal} {Physics Reports}\ }\textbf {\bibinfo {volume} {492}},\ \bibinfo {pages} {1 } (\bibinfo {year} {2010})}\BibitemShut {NoStop}%
\bibitem [{\citenamefont {Shevchenko}\ \emph {et~al.}(2018)\citenamefont {Shevchenko}, \citenamefont {Ryzhov},\ and\ \citenamefont {Nori}}]{shevchenko_2018}%
  \BibitemOpen
  \bibfield  {author} {\bibinfo {author} {\bibfnamefont {S.~N.}\ \bibnamefont {Shevchenko}}, \bibinfo {author} {\bibfnamefont {A.~I.}\ \bibnamefont {Ryzhov}},\ and\ \bibinfo {author} {\bibfnamefont {F.}~\bibnamefont {Nori}},\ }\bibfield  {title} {\bibinfo {title} {Low-frequency spectroscopy for quantum multilevel systems},\ }\href {https://doi.org/10.1103/PhysRevB.98.195434} {\bibfield  {journal} {\bibinfo  {journal} {Phys. Rev. B}\ }\textbf {\bibinfo {volume} {98}},\ \bibinfo {pages} {195434} (\bibinfo {year} {2018})}\BibitemShut {NoStop}%
\bibitem [{\citenamefont {Ivakhnenko}\ \emph {et~al.}(2023)\citenamefont {Ivakhnenko}, \citenamefont {Shevchenko},\ and\ \citenamefont {Nori}}]{ivakhnenko_2023}%
  \BibitemOpen
  \bibfield  {author} {\bibinfo {author} {\bibfnamefont {O.~V.}\ \bibnamefont {Ivakhnenko}}, \bibinfo {author} {\bibfnamefont {S.~N.}\ \bibnamefont {Shevchenko}},\ and\ \bibinfo {author} {\bibfnamefont {F.}~\bibnamefont {Nori}},\ }\bibfield  {title} {\bibinfo {title} {Nonadiabatic landau–zener–stückelberg–majorana transitions, dynamics, and interference},\ }\href {https://doi.org/https://doi.org/10.1016/j.physrep.2022.10.002} {\bibfield  {journal} {\bibinfo  {journal} {Physics Reports}\ }\textbf {\bibinfo {volume} {995}},\ \bibinfo {pages} {1} (\bibinfo {year} {2023})},\ \bibinfo {note} {nonadiabatic Landau-Zener-Stückelberg-Majorana transitions, dynamics, and interference}\BibitemShut {NoStop}%
\bibitem [{\citenamefont {Oliver}\ \emph {et~al.}(2005)\citenamefont {Oliver}, \citenamefont {Yu}, \citenamefont {Lee}, \citenamefont {Berggren}, \citenamefont {Levitov},\ and\ \citenamefont {Orlando}}]{Oliver_2005}%
  \BibitemOpen
  \bibfield  {author} {\bibinfo {author} {\bibfnamefont {W.~D.}\ \bibnamefont {Oliver}}, \bibinfo {author} {\bibfnamefont {Y.}~\bibnamefont {Yu}}, \bibinfo {author} {\bibfnamefont {J.~C.}\ \bibnamefont {Lee}}, \bibinfo {author} {\bibfnamefont {K.~K.}\ \bibnamefont {Berggren}}, \bibinfo {author} {\bibfnamefont {L.~S.}\ \bibnamefont {Levitov}},\ and\ \bibinfo {author} {\bibfnamefont {T.~P.}\ \bibnamefont {Orlando}},\ }\bibfield  {title} {\bibinfo {title} {Mach-zehnder interferometry in a strongly driven superconducting qubit},\ }\href@noop {} {\bibfield  {journal} {\bibinfo  {journal} {Science}\ }\textbf {\bibinfo {volume} {310}},\ \bibinfo {pages} {1653} (\bibinfo {year} {2005})}\BibitemShut {NoStop}%
\bibitem [{\citenamefont {Sillanp\"a\"a}\ \emph {et~al.}(2006)\citenamefont {Sillanp\"a\"a}, \citenamefont {Lehtinen}, \citenamefont {Paila}, \citenamefont {Makhlin},\ and\ \citenamefont {Hakonen}}]{Sillanpaa_2006}%
  \BibitemOpen
  \bibfield  {author} {\bibinfo {author} {\bibfnamefont {M.}~\bibnamefont {Sillanp\"a\"a}}, \bibinfo {author} {\bibfnamefont {T.}~\bibnamefont {Lehtinen}}, \bibinfo {author} {\bibfnamefont {A.}~\bibnamefont {Paila}}, \bibinfo {author} {\bibfnamefont {Y.}~\bibnamefont {Makhlin}},\ and\ \bibinfo {author} {\bibfnamefont {P.}~\bibnamefont {Hakonen}},\ }\bibfield  {title} {\bibinfo {title} {Continuous-time monitoring of landau-zener interference in a cooper-pair box},\ }\href {https://doi.org/10.1103/PhysRevLett.96.187002} {\bibfield  {journal} {\bibinfo  {journal} {Phys. Rev. Lett.}\ }\textbf {\bibinfo {volume} {96}},\ \bibinfo {pages} {187002} (\bibinfo {year} {2006})}\BibitemShut {NoStop}%
\bibitem [{\citenamefont {Izmalkov}\ \emph {et~al.}(2008)\citenamefont {Izmalkov}, \citenamefont {van~der Ploeg}, \citenamefont {Shevchenko}, \citenamefont {Grajcar}, \citenamefont {Il'ichev}, \citenamefont {H\"ubner}, \citenamefont {Omelyanchouk},\ and\ \citenamefont {Meyer}}]{Izmalkov_2008}%
  \BibitemOpen
  \bibfield  {author} {\bibinfo {author} {\bibfnamefont {A.}~\bibnamefont {Izmalkov}}, \bibinfo {author} {\bibfnamefont {S.~H.~W.}\ \bibnamefont {van~der Ploeg}}, \bibinfo {author} {\bibfnamefont {S.~N.}\ \bibnamefont {Shevchenko}}, \bibinfo {author} {\bibfnamefont {M.}~\bibnamefont {Grajcar}}, \bibinfo {author} {\bibfnamefont {E.}~\bibnamefont {Il'ichev}}, \bibinfo {author} {\bibfnamefont {U.}~\bibnamefont {H\"ubner}}, \bibinfo {author} {\bibfnamefont {A.~N.}\ \bibnamefont {Omelyanchouk}},\ and\ \bibinfo {author} {\bibfnamefont {H.-G.}\ \bibnamefont {Meyer}},\ }\bibfield  {title} {\bibinfo {title} {Consistency of ground state and spectroscopic measurements on flux qubits},\ }\href {https://doi.org/10.1103/PhysRevLett.101.017003} {\bibfield  {journal} {\bibinfo  {journal} {Phys. Rev. Lett.}\ }\textbf {\bibinfo {volume} {101}},\ \bibinfo {pages} {017003} (\bibinfo {year} {2008})}\BibitemShut {NoStop}%
\bibitem [{\citenamefont {Oliver}\ and\ \citenamefont {Valenzuela}(2009)}]{Oliver_2009}%
  \BibitemOpen
  \bibfield  {author} {\bibinfo {author} {\bibfnamefont {W.~D.}\ \bibnamefont {Oliver}}\ and\ \bibinfo {author} {\bibfnamefont {S.~O.}\ \bibnamefont {Valenzuela}},\ }\bibfield  {title} {\bibinfo {title} {Large-amplitude driving of a superconducting artificial atom},\ }\href@noop {} {\bibfield  {journal} {\bibinfo  {journal} {Quantum Information Processing}\ }\textbf {\bibinfo {volume} {8}},\ \bibinfo {pages} {261} (\bibinfo {year} {2009})}\BibitemShut {NoStop}%
\bibitem [{\citenamefont {Berns}\ \emph {et~al.}(2008)\citenamefont {Berns}, \citenamefont {Rudner}, \citenamefont {Valenzuela}, \citenamefont {Berggren}, \citenamefont {Oliver}, \citenamefont {Levitov},\ and\ \citenamefont {Orlando}}]{Berns_2008}%
  \BibitemOpen
  \bibfield  {author} {\bibinfo {author} {\bibfnamefont {D.~M.}\ \bibnamefont {Berns}}, \bibinfo {author} {\bibfnamefont {M.~S.}\ \bibnamefont {Rudner}}, \bibinfo {author} {\bibfnamefont {S.~O.}\ \bibnamefont {Valenzuela}}, \bibinfo {author} {\bibfnamefont {K.~K.}\ \bibnamefont {Berggren}}, \bibinfo {author} {\bibfnamefont {W.~D.}\ \bibnamefont {Oliver}}, \bibinfo {author} {\bibfnamefont {L.~S.}\ \bibnamefont {Levitov}},\ and\ \bibinfo {author} {\bibfnamefont {T.~P.}\ \bibnamefont {Orlando}},\ }\bibfield  {title} {\bibinfo {title} {Amplitude spectroscopy of a solid-state artificial atom},\ }\href@noop {} {\bibfield  {journal} {\bibinfo  {journal} {Nature}\ }\textbf {\bibinfo {volume} {455}},\ \bibinfo {pages} {51} (\bibinfo {year} {2008})}\BibitemShut {NoStop}%
\bibitem [{\citenamefont {Ferr\'on}\ \emph {et~al.}(2012)\citenamefont {Ferr\'on}, \citenamefont {Dom\'{\i}nguez},\ and\ \citenamefont {S\'anchez}}]{Ferron_2012}%
  \BibitemOpen
  \bibfield  {author} {\bibinfo {author} {\bibfnamefont {A.}~\bibnamefont {Ferr\'on}}, \bibinfo {author} {\bibfnamefont {D.}~\bibnamefont {Dom\'{\i}nguez}},\ and\ \bibinfo {author} {\bibfnamefont {M.~J.}\ \bibnamefont {S\'anchez}},\ }\bibfield  {title} {\bibinfo {title} {Tailoring population inversion in landau-zener-st\"uckelberg interferometry of flux qubits},\ }\href {https://doi.org/10.1103/PhysRevLett.109.237005} {\bibfield  {journal} {\bibinfo  {journal} {Phys. Rev. Lett.}\ }\textbf {\bibinfo {volume} {109}},\ \bibinfo {pages} {237005} (\bibinfo {year} {2012})}\BibitemShut {NoStop}%
\bibitem [{\citenamefont {Gustavsson}\ \emph {et~al.}(2012)\citenamefont {Gustavsson}, \citenamefont {Bylander}, \citenamefont {Yan}, \citenamefont {Forn-D\'{\i}az}, \citenamefont {Bolkhovsky}, \citenamefont {Braje}, \citenamefont {Fitch}, \citenamefont {Harrabi}, \citenamefont {Lennon}, \citenamefont {Miloshi}, \citenamefont {Murphy}, \citenamefont {Slattery}, \citenamefont {Spector}, \citenamefont {Turek}, \citenamefont {Weir}, \citenamefont {Welander}, \citenamefont {Yoshihara}, \citenamefont {Cory}, \citenamefont {Nakamura}, \citenamefont {Orlando},\ and\ \citenamefont {Oliver}}]{gustavsson_2012}%
  \BibitemOpen
  \bibfield  {author} {\bibinfo {author} {\bibfnamefont {S.}~\bibnamefont {Gustavsson}}, \bibinfo {author} {\bibfnamefont {J.}~\bibnamefont {Bylander}}, \bibinfo {author} {\bibfnamefont {F.}~\bibnamefont {Yan}}, \bibinfo {author} {\bibfnamefont {P.}~\bibnamefont {Forn-D\'{\i}az}}, \bibinfo {author} {\bibfnamefont {V.}~\bibnamefont {Bolkhovsky}}, \bibinfo {author} {\bibfnamefont {D.}~\bibnamefont {Braje}}, \bibinfo {author} {\bibfnamefont {G.}~\bibnamefont {Fitch}}, \bibinfo {author} {\bibfnamefont {K.}~\bibnamefont {Harrabi}}, \bibinfo {author} {\bibfnamefont {D.}~\bibnamefont {Lennon}}, \bibinfo {author} {\bibfnamefont {J.}~\bibnamefont {Miloshi}}, \bibinfo {author} {\bibfnamefont {P.}~\bibnamefont {Murphy}}, \bibinfo {author} {\bibfnamefont {R.}~\bibnamefont {Slattery}}, \bibinfo {author} {\bibfnamefont {S.}~\bibnamefont {Spector}}, \bibinfo {author} {\bibfnamefont {B.}~\bibnamefont {Turek}}, \bibinfo {author} {\bibfnamefont {T.}~\bibnamefont {Weir}}, \bibinfo {author} {\bibfnamefont {P.~B.}\ \bibnamefont
  {Welander}}, \bibinfo {author} {\bibfnamefont {F.}~\bibnamefont {Yoshihara}}, \bibinfo {author} {\bibfnamefont {D.~G.}\ \bibnamefont {Cory}}, \bibinfo {author} {\bibfnamefont {Y.}~\bibnamefont {Nakamura}}, \bibinfo {author} {\bibfnamefont {T.~P.}\ \bibnamefont {Orlando}},\ and\ \bibinfo {author} {\bibfnamefont {W.~D.}\ \bibnamefont {Oliver}},\ }\bibfield  {title} {\bibinfo {title} {Driven dynamics and rotary echo of a qubit tunably coupled to a harmonic oscillator},\ }\href {https://doi.org/10.1103/PhysRevLett.108.170503} {\bibfield  {journal} {\bibinfo  {journal} {Phys. Rev. Lett.}\ }\textbf {\bibinfo {volume} {108}},\ \bibinfo {pages} {170503} (\bibinfo {year} {2012})}\BibitemShut {NoStop}%
\bibitem [{\citenamefont {Ferr\'on}\ \emph {et~al.}(2016)\citenamefont {Ferr\'on}, \citenamefont {Dom\'{\i}nguez},\ and\ \citenamefont {S\'anchez}}]{Ferron_2016}%
  \BibitemOpen
  \bibfield  {author} {\bibinfo {author} {\bibfnamefont {A.}~\bibnamefont {Ferr\'on}}, \bibinfo {author} {\bibfnamefont {D.}~\bibnamefont {Dom\'{\i}nguez}},\ and\ \bibinfo {author} {\bibfnamefont {M.~J.}\ \bibnamefont {S\'anchez}},\ }\bibfield  {title} {\bibinfo {title} {Dynamic transition in landau-zener-st\"uckelberg interferometry of dissipative systems: The case of the flux qubit},\ }\href {https://doi.org/10.1103/PhysRevB.93.064521} {\bibfield  {journal} {\bibinfo  {journal} {Phys. Rev. B}\ }\textbf {\bibinfo {volume} {93}},\ \bibinfo {pages} {064521} (\bibinfo {year} {2016})}\BibitemShut {NoStop}%
\bibitem [{\citenamefont {Gramajo}\ \emph {et~al.}(2017)\citenamefont {Gramajo}, \citenamefont {Domínguez},\ and\ \citenamefont {S\'anchez}}]{Gramajo_2017}%
  \BibitemOpen
  \bibfield  {author} {\bibinfo {author} {\bibfnamefont {A.~L.}\ \bibnamefont {Gramajo}}, \bibinfo {author} {\bibfnamefont {D.}~\bibnamefont {Domínguez}},\ and\ \bibinfo {author} {\bibfnamefont {M.~J.}\ \bibnamefont {S\'anchez}},\ }\bibfield  {title} {\bibinfo {title} {Entanglement generation through the interplay of harmonic driving and interaction in coupled superconducting qubits},\ }\href@noop {} {\bibfield  {journal} {\bibinfo  {journal} {Eur. Phys. J. B}\ }\textbf {\bibinfo {volume} {90}},\ \bibinfo {pages} {255} (\bibinfo {year} {2017})}\BibitemShut {NoStop}%
\bibitem [{\citenamefont {Gramajo}\ \emph {et~al.}(2018)\citenamefont {Gramajo}, \citenamefont {Dom\'{\i}nguez},\ and\ \citenamefont {S\'anchez}}]{gramajo_2018}%
  \BibitemOpen
  \bibfield  {author} {\bibinfo {author} {\bibfnamefont {A.~L.}\ \bibnamefont {Gramajo}}, \bibinfo {author} {\bibfnamefont {D.}~\bibnamefont {Dom\'{\i}nguez}},\ and\ \bibinfo {author} {\bibfnamefont {M.~J.}\ \bibnamefont {S\'anchez}},\ }\bibfield  {title} {\bibinfo {title} {Amplitude tuning of steady-state entanglement in strongly driven coupled qubits},\ }\href {https://doi.org/10.1103/PhysRevA.98.042337} {\bibfield  {journal} {\bibinfo  {journal} {Phys. Rev. A}\ }\textbf {\bibinfo {volume} {98}},\ \bibinfo {pages} {042337} (\bibinfo {year} {2018})}\BibitemShut {NoStop}%
\bibitem [{\citenamefont {Gramajo}\ \emph {et~al.}(2021)\citenamefont {Gramajo}, \citenamefont {Dom\'{\i}nguez},\ and\ \citenamefont {S\'anchez}}]{gramajo_2021}%
  \BibitemOpen
  \bibfield  {author} {\bibinfo {author} {\bibfnamefont {A.~L.}\ \bibnamefont {Gramajo}}, \bibinfo {author} {\bibfnamefont {D.}~\bibnamefont {Dom\'{\i}nguez}},\ and\ \bibinfo {author} {\bibfnamefont {M.~J.}\ \bibnamefont {S\'anchez}},\ }\bibfield  {title} {\bibinfo {title} {Efficient steady-state-entanglement generation in strongly driven coupled qubits},\ }\href {https://doi.org/10.1103/PhysRevA.104.032410} {\bibfield  {journal} {\bibinfo  {journal} {Phys. Rev. A}\ }\textbf {\bibinfo {volume} {104}},\ \bibinfo {pages} {032410} (\bibinfo {year} {2021})}\BibitemShut {NoStop}%
\bibitem [{\citenamefont {Bonifacio}\ \emph {et~al.}(2020)\citenamefont {Bonifacio}, \citenamefont {Dom{\'\i}nguez},\ and\ \citenamefont {S{\'a}nchez}}]{Bonifacio_2020}%
  \BibitemOpen
  \bibfield  {author} {\bibinfo {author} {\bibfnamefont {M.}~\bibnamefont {Bonifacio}}, \bibinfo {author} {\bibfnamefont {D.}~\bibnamefont {Dom{\'\i}nguez}},\ and\ \bibinfo {author} {\bibfnamefont {M.~J.}\ \bibnamefont {S{\'a}nchez}},\ }\bibfield  {title} {\bibinfo {title} {Landau-zener-st{\"u}ckelberg interferometry in dissipative circuit quantum electrodynamics},\ }\href@noop {} {\bibfield  {journal} {\bibinfo  {journal} {Physical Review B}\ }\textbf {\bibinfo {volume} {101}},\ \bibinfo {pages} {245415} (\bibinfo {year} {2020})}\BibitemShut {NoStop}%
\bibitem [{\citenamefont {Gallardo}\ \emph {et~al.}(2022)\citenamefont {Gallardo}, \citenamefont {Dom\'{\i}nguez},\ and\ \citenamefont {S\'anchez}}]{gallardo_2022}%
  \BibitemOpen
  \bibfield  {author} {\bibinfo {author} {\bibfnamefont {S.~L.}\ \bibnamefont {Gallardo}}, \bibinfo {author} {\bibfnamefont {D.}~\bibnamefont {Dom\'{\i}nguez}},\ and\ \bibinfo {author} {\bibfnamefont {M.~J.}\ \bibnamefont {S\'anchez}},\ }\bibfield  {title} {\bibinfo {title} {Dissipative entanglement generation between two qubits parametrically driven and coupled to a resonator},\ }\href {https://doi.org/10.1103/PhysRevA.105.052413} {\bibfield  {journal} {\bibinfo  {journal} {Phys. Rev. A}\ }\textbf {\bibinfo {volume} {105}},\ \bibinfo {pages} {052413} (\bibinfo {year} {2022})}\BibitemShut {NoStop}%
\bibitem [{\citenamefont {Campbell}\ \emph {et~al.}(2020)\citenamefont {Campbell}, \citenamefont {Shim}, \citenamefont {Kannan}, \citenamefont {Winik}, \citenamefont {Kim}, \citenamefont {Melville}, \citenamefont {Niedzielski}, \citenamefont {Yoder}, \citenamefont {Tahan}, \citenamefont {Gustavsson},\ and\ \citenamefont {Oliver}}]{campbell_2020}%
  \BibitemOpen
  \bibfield  {author} {\bibinfo {author} {\bibfnamefont {D.~L.}\ \bibnamefont {Campbell}}, \bibinfo {author} {\bibfnamefont {Y.-P.}\ \bibnamefont {Shim}}, \bibinfo {author} {\bibfnamefont {B.}~\bibnamefont {Kannan}}, \bibinfo {author} {\bibfnamefont {R.}~\bibnamefont {Winik}}, \bibinfo {author} {\bibfnamefont {D.~K.}\ \bibnamefont {Kim}}, \bibinfo {author} {\bibfnamefont {A.}~\bibnamefont {Melville}}, \bibinfo {author} {\bibfnamefont {B.~M.}\ \bibnamefont {Niedzielski}}, \bibinfo {author} {\bibfnamefont {J.~L.}\ \bibnamefont {Yoder}}, \bibinfo {author} {\bibfnamefont {C.}~\bibnamefont {Tahan}}, \bibinfo {author} {\bibfnamefont {S.}~\bibnamefont {Gustavsson}},\ and\ \bibinfo {author} {\bibfnamefont {W.~D.}\ \bibnamefont {Oliver}},\ }\bibfield  {title} {\bibinfo {title} {Universal nonadiabatic control of small-gap superconducting qubits},\ }\href {https://doi.org/10.1103/PhysRevX.10.041051} {\bibfield  {journal} {\bibinfo  {journal} {Phys. Rev. X}\ }\textbf {\bibinfo {volume} {10}},\ \bibinfo {pages} {041051}
  (\bibinfo {year} {2020})}\BibitemShut {NoStop}%
\bibitem [{\citenamefont {C\'aceres}\ \emph {et~al.}(2023)\citenamefont {C\'aceres}, \citenamefont {Dom\'{\i}nguez},\ and\ \citenamefont {S\'anchez}}]{caceres_2023}%
  \BibitemOpen
  \bibfield  {author} {\bibinfo {author} {\bibfnamefont {J.~J.}\ \bibnamefont {C\'aceres}}, \bibinfo {author} {\bibfnamefont {D.}~\bibnamefont {Dom\'{\i}nguez}},\ and\ \bibinfo {author} {\bibfnamefont {M.~J.}\ \bibnamefont {S\'anchez}},\ }\bibfield  {title} {\bibinfo {title} {Fast quantum gates based on landau-zener-st\"uckelberg-majorana transitions},\ }\href {https://doi.org/10.1103/PhysRevA.108.052619} {\bibfield  {journal} {\bibinfo  {journal} {Phys. Rev. A}\ }\textbf {\bibinfo {volume} {108}},\ \bibinfo {pages} {052619} (\bibinfo {year} {2023})}\BibitemShut {NoStop}%
\bibitem [{\citenamefont {Mundada}\ \emph {et~al.}(2020)\citenamefont {Mundada}, \citenamefont {Gyenis}, \citenamefont {Huang}, \citenamefont {Koch},\ and\ \citenamefont {Houck}}]{mundada_2020}%
  \BibitemOpen
  \bibfield  {author} {\bibinfo {author} {\bibfnamefont {P.~S.}\ \bibnamefont {Mundada}}, \bibinfo {author} {\bibfnamefont {A.}~\bibnamefont {Gyenis}}, \bibinfo {author} {\bibfnamefont {Z.}~\bibnamefont {Huang}}, \bibinfo {author} {\bibfnamefont {J.}~\bibnamefont {Koch}},\ and\ \bibinfo {author} {\bibfnamefont {A.~A.}\ \bibnamefont {Houck}},\ }\bibfield  {title} {\bibinfo {title} {Floquet-engineered enhancement of coherence times in a driven fluxonium qubit},\ }\href {https://doi.org/10.1103/PhysRevApplied.14.054033} {\bibfield  {journal} {\bibinfo  {journal} {Phys. Rev. Appl.}\ }\textbf {\bibinfo {volume} {14}},\ \bibinfo {pages} {054033} (\bibinfo {year} {2020})}\BibitemShut {NoStop}%
\bibitem [{\citenamefont {Schrieffer}\ and\ \citenamefont {Wolff}(1966)}]{schrieffer1966relation}%
  \BibitemOpen
  \bibfield  {author} {\bibinfo {author} {\bibfnamefont {J.~R.}\ \bibnamefont {Schrieffer}}\ and\ \bibinfo {author} {\bibfnamefont {P.~A.}\ \bibnamefont {Wolff}},\ }\bibfield  {title} {\bibinfo {title} {Relation between the anderson and kondo hamiltonians},\ }\href@noop {} {\bibfield  {journal} {\bibinfo  {journal} {Physical Review}\ }\textbf {\bibinfo {volume} {149}},\ \bibinfo {pages} {491} (\bibinfo {year} {1966})}\BibitemShut {NoStop}%
\bibitem [{\citenamefont {Bravyi}\ \emph {et~al.}(2011)\citenamefont {Bravyi}, \citenamefont {DiVincenzo},\ and\ \citenamefont {Loss}}]{bravyi2011schrieffer}%
  \BibitemOpen
  \bibfield  {author} {\bibinfo {author} {\bibfnamefont {S.}~\bibnamefont {Bravyi}}, \bibinfo {author} {\bibfnamefont {D.~P.}\ \bibnamefont {DiVincenzo}},\ and\ \bibinfo {author} {\bibfnamefont {D.}~\bibnamefont {Loss}},\ }\bibfield  {title} {\bibinfo {title} {Schrieffer--wolff transformation for quantum many-body systems},\ }\href@noop {} {\bibfield  {journal} {\bibinfo  {journal} {Annals of physics}\ }\textbf {\bibinfo {volume} {326}},\ \bibinfo {pages} {2793} (\bibinfo {year} {2011})}\BibitemShut {NoStop}%
\bibitem [{\citenamefont {Santoro}(2019)}]{santoro2019introduction}%
  \BibitemOpen
  \bibfield  {author} {\bibinfo {author} {\bibfnamefont {G.~E.}\ \bibnamefont {Santoro}},\ }\bibfield  {title} {\bibinfo {title} {Introduction to floquet},\ }\href@noop {} {\bibfield  {journal} {\bibinfo  {journal} {Lecture Notes}\ } (\bibinfo {year} {2019})}\BibitemShut {NoStop}%
\bibitem [{\citenamefont {Bukov}\ \emph {et~al.}(2015)\citenamefont {Bukov}, \citenamefont {D'Alessio},\ and\ \citenamefont {Polkovnikov}}]{bukov2015universal}%
  \BibitemOpen
  \bibfield  {author} {\bibinfo {author} {\bibfnamefont {M.}~\bibnamefont {Bukov}}, \bibinfo {author} {\bibfnamefont {L.}~\bibnamefont {D'Alessio}},\ and\ \bibinfo {author} {\bibfnamefont {A.}~\bibnamefont {Polkovnikov}},\ }\bibfield  {title} {\bibinfo {title} {Universal high-frequency behavior of periodically driven systems: from dynamical stabilization to floquet engineering},\ }\href@noop {} {\bibfield  {journal} {\bibinfo  {journal} {Advances in Physics}\ }\textbf {\bibinfo {volume} {64}},\ \bibinfo {pages} {139} (\bibinfo {year} {2015})}\BibitemShut {NoStop}%
\bibitem [{\citenamefont {Johansson}\ \emph {et~al.}(2012)\citenamefont {Johansson}, \citenamefont {Nation},\ and\ \citenamefont {Nori}}]{johansson2012qutip}%
  \BibitemOpen
  \bibfield  {author} {\bibinfo {author} {\bibfnamefont {J.~R.}\ \bibnamefont {Johansson}}, \bibinfo {author} {\bibfnamefont {P.~D.}\ \bibnamefont {Nation}},\ and\ \bibinfo {author} {\bibfnamefont {F.}~\bibnamefont {Nori}},\ }\bibfield  {title} {\bibinfo {title} {Qutip: An open-source python framework for the dynamics of open quantum systems},\ }\href@noop {} {\bibfield  {journal} {\bibinfo  {journal} {Computer physics communications}\ }\textbf {\bibinfo {volume} {183}},\ \bibinfo {pages} {1760} (\bibinfo {year} {2012})}\BibitemShut {NoStop}%
\bibitem [{\citenamefont {Stehlik}\ \emph {et~al.}(2012)\citenamefont {Stehlik}, \citenamefont {Dovzhenko}, \citenamefont {Petta}, \citenamefont {Johansson}, \citenamefont {Nori}, \citenamefont {Lu},\ and\ \citenamefont {Gossard}}]{stehlik_2012}%
  \BibitemOpen
  \bibfield  {author} {\bibinfo {author} {\bibfnamefont {J.}~\bibnamefont {Stehlik}}, \bibinfo {author} {\bibfnamefont {Y.}~\bibnamefont {Dovzhenko}}, \bibinfo {author} {\bibfnamefont {J.~R.}\ \bibnamefont {Petta}}, \bibinfo {author} {\bibfnamefont {J.~R.}\ \bibnamefont {Johansson}}, \bibinfo {author} {\bibfnamefont {F.}~\bibnamefont {Nori}}, \bibinfo {author} {\bibfnamefont {H.}~\bibnamefont {Lu}},\ and\ \bibinfo {author} {\bibfnamefont {A.~C.}\ \bibnamefont {Gossard}},\ }\bibfield  {title} {\bibinfo {title} {Landau-zener-st\"uckelberg interferometry of a single electron charge qubit},\ }\href {https://doi.org/10.1103/PhysRevB.86.121303} {\bibfield  {journal} {\bibinfo  {journal} {Phys. Rev. B}\ }\textbf {\bibinfo {volume} {86}},\ \bibinfo {pages} {121303} (\bibinfo {year} {2012})}\BibitemShut {NoStop}%
\bibitem [{\citenamefont {Koski}\ \emph {et~al.}(2018)\citenamefont {Koski}, \citenamefont {Landig}, \citenamefont {P\'alyi}, \citenamefont {Scarlino}, \citenamefont {Reichl}, \citenamefont {Wegscheider}, \citenamefont {Burkard}, \citenamefont {Wallraff}, \citenamefont {Ensslin},\ and\ \citenamefont {Ihn}}]{koski_2018}%
  \BibitemOpen
  \bibfield  {author} {\bibinfo {author} {\bibfnamefont {J.~V.}\ \bibnamefont {Koski}}, \bibinfo {author} {\bibfnamefont {A.~J.}\ \bibnamefont {Landig}}, \bibinfo {author} {\bibfnamefont {A.}~\bibnamefont {P\'alyi}}, \bibinfo {author} {\bibfnamefont {P.}~\bibnamefont {Scarlino}}, \bibinfo {author} {\bibfnamefont {C.}~\bibnamefont {Reichl}}, \bibinfo {author} {\bibfnamefont {W.}~\bibnamefont {Wegscheider}}, \bibinfo {author} {\bibfnamefont {G.}~\bibnamefont {Burkard}}, \bibinfo {author} {\bibfnamefont {A.}~\bibnamefont {Wallraff}}, \bibinfo {author} {\bibfnamefont {K.}~\bibnamefont {Ensslin}},\ and\ \bibinfo {author} {\bibfnamefont {T.}~\bibnamefont {Ihn}},\ }\bibfield  {title} {\bibinfo {title} {Floquet spectroscopy of a strongly driven quantum dot charge qubit with a microwave resonator},\ }\href {https://doi.org/10.1103/PhysRevLett.121.043603} {\bibfield  {journal} {\bibinfo  {journal} {Phys. Rev. Lett.}\ }\textbf {\bibinfo {volume} {121}},\ \bibinfo {pages} {043603} (\bibinfo {year} {2018})}\BibitemShut
  {NoStop}%
\bibitem [{\citenamefont {Gramajo}\ \emph {et~al.}(2020)\citenamefont {Gramajo}, \citenamefont {Campbell}, \citenamefont {Kannan}, \citenamefont {Kim}, \citenamefont {Melville}, \citenamefont {Niedzielski}, \citenamefont {Yoder}, \citenamefont {S\'anchez}, \citenamefont {Dom\'{\i}nguez}, \citenamefont {Gustavsson},\ and\ \citenamefont {Oliver}}]{gramajo_2020}%
  \BibitemOpen
  \bibfield  {author} {\bibinfo {author} {\bibfnamefont {A.~L.}\ \bibnamefont {Gramajo}}, \bibinfo {author} {\bibfnamefont {D.}~\bibnamefont {Campbell}}, \bibinfo {author} {\bibfnamefont {B.}~\bibnamefont {Kannan}}, \bibinfo {author} {\bibfnamefont {D.~K.}\ \bibnamefont {Kim}}, \bibinfo {author} {\bibfnamefont {A.}~\bibnamefont {Melville}}, \bibinfo {author} {\bibfnamefont {B.~M.}\ \bibnamefont {Niedzielski}}, \bibinfo {author} {\bibfnamefont {J.~L.}\ \bibnamefont {Yoder}}, \bibinfo {author} {\bibfnamefont {M.~J.}\ \bibnamefont {S\'anchez}}, \bibinfo {author} {\bibfnamefont {D.}~\bibnamefont {Dom\'{\i}nguez}}, \bibinfo {author} {\bibfnamefont {S.}~\bibnamefont {Gustavsson}},\ and\ \bibinfo {author} {\bibfnamefont {W.~D.}\ \bibnamefont {Oliver}},\ }\bibfield  {title} {\bibinfo {title} {Quantum emulation of coherent backscattering in a system of superconducting qubits},\ }\href {https://doi.org/10.1103/PhysRevApplied.14.014047} {\bibfield  {journal} {\bibinfo  {journal} {Phys. Rev. Applied}\ }\textbf {\bibinfo
  {volume} {14}},\ \bibinfo {pages} {014047} (\bibinfo {year} {2020})}\BibitemShut {NoStop}%
\bibitem [{Note1()}]{Note1}%
  \BibitemOpen
  \bibinfo {note} {At a first glance, the two-level truncation seems to be a quite reasonable approach at the operationally ``sweet spot'', $\delta _0=\pi $, since $E_2(\pi )-E_1(\pi ) \gg E_1(\pi )-E_0(\pi )$. For the sake of comparison, the qubit Hamiltonian at $\delta _0=\pi $ is $\protect \hat {H}\approx \protect \frac {\omega _q(\pi )}2 \sigma _z - 2 E_L A \sigma _x \cos (\omega t)$, such that the Rabi frequency for a resonant tone ($\omega =\omega _q(\pi )$) is $\Omega _R\approx 4 E_L A$. This means that the amplitude to realize a PI-pulse in 10\ ns would be $A\approx 0.004 \pi $ ($\Omega _R=2\pi \times 0.05\ $GHz). As we are considering non-resonant tones and much larger amplitudes, the full LZS spectrum (see Fig. \ref {fig:fig3}) differs significantly from what is shown in Fig. \ref {fig:fig2}}\BibitemShut {NoStop}%
\bibitem [{\citenamefont {Shirley}(1965)}]{shirley_1965}%
  \BibitemOpen
  \bibfield  {author} {\bibinfo {author} {\bibfnamefont {J.~H.}\ \bibnamefont {Shirley}},\ }\bibfield  {title} {\bibinfo {title} {Solution of the schr{\"o}dinger equation with a hamiltonian periodic in time},\ }\href@noop {} {\bibfield  {journal} {\bibinfo  {journal} {Physical Review}\ }\textbf {\bibinfo {volume} {138}},\ \bibinfo {pages} {B979} (\bibinfo {year} {1965})}\BibitemShut {NoStop}%
\bibitem [{\citenamefont {Kohler}\ \emph {et~al.}(1997)\citenamefont {Kohler}, \citenamefont {Dittrich},\ and\ \citenamefont {H\"anggi}}]{kohler_1997}%
  \BibitemOpen
  \bibfield  {author} {\bibinfo {author} {\bibfnamefont {S.}~\bibnamefont {Kohler}}, \bibinfo {author} {\bibfnamefont {T.}~\bibnamefont {Dittrich}},\ and\ \bibinfo {author} {\bibfnamefont {P.}~\bibnamefont {H\"anggi}},\ }\bibfield  {title} {\bibinfo {title} {Floquet-markovian description of the parametrically driven, dissipative harmonic quantum oscillator},\ }\href {https://doi.org/10.1103/PhysRevE.55.300} {\bibfield  {journal} {\bibinfo  {journal} {Phys. Rev. E}\ }\textbf {\bibinfo {volume} {55}},\ \bibinfo {pages} {300} (\bibinfo {year} {1997})}\BibitemShut {NoStop}%
\bibitem [{\citenamefont {Son}\ \emph {et~al.}(2009)\citenamefont {Son}, \citenamefont {Han},\ and\ \citenamefont {Chu}}]{son_2009}%
  \BibitemOpen
  \bibfield  {author} {\bibinfo {author} {\bibfnamefont {S.-K.}\ \bibnamefont {Son}}, \bibinfo {author} {\bibfnamefont {S.}~\bibnamefont {Han}},\ and\ \bibinfo {author} {\bibfnamefont {S.-I.}\ \bibnamefont {Chu}},\ }\bibfield  {title} {\bibinfo {title} {Floquet formulation for the investigation of multiphoton quantum interference in a superconducting qubit driven by a strong ac field},\ }\href {https://doi.org/10.1103/PhysRevA.79.032301} {\bibfield  {journal} {\bibinfo  {journal} {Phys. Rev. A}\ }\textbf {\bibinfo {volume} {79}},\ \bibinfo {pages} {032301} (\bibinfo {year} {2009})}\BibitemShut {NoStop}%
\bibitem [{\citenamefont {Grifoni}\ and\ \citenamefont {H{\"a}nggi}(1998)}]{grifoni_1998}%
  \BibitemOpen
  \bibfield  {author} {\bibinfo {author} {\bibfnamefont {M.}~\bibnamefont {Grifoni}}\ and\ \bibinfo {author} {\bibfnamefont {P.}~\bibnamefont {H{\"a}nggi}},\ }\bibfield  {title} {\bibinfo {title} {Driven quantum tunneling},\ }\href {https://doi.org/https://doi.org/10.1016/S0370-1573(98)00022-2} {\bibfield  {journal} {\bibinfo  {journal} {Physics Reports}\ }\textbf {\bibinfo {volume} {304}},\ \bibinfo {pages} {229 } (\bibinfo {year} {1998})}\BibitemShut {NoStop}%
\end{thebibliography}%
\end{document}